\documentclass[prl,twocolumn,amsmath,amssymb,floatfix,reprint,footinbib,superscriptaddress,longbibliography,showkeys]{revtex4-1}
\usepackage{dcolumn}
\usepackage{graphicx}
\usepackage{mathrsfs}
\usepackage{mdwlist}
\usepackage{subfigure}
\usepackage{booktabs}
\usepackage{multirow}
\usepackage{amsmath}
\usepackage{textcomp}
\usepackage{upgreek}
\usepackage{dsfont}
\usepackage{amstext}
\usepackage{amssymb}
\usepackage{amsbsy}
\usepackage{appendix}
\usepackage{soul}
\usepackage{threeparttable}
\usepackage{bbm}
\usepackage{amsthm}
\usepackage{graphicx}
\usepackage{lipsum}
\usepackage{xcolor}
\usepackage[none]{hyphenat}
\usepackage{textcomp}
\usepackage{color}
\usepackage[colorlinks,citecolor=blue]{hyperref}
\definecolor{Dgreen}{RGB}{0, 100, 0}
\usepackage{url}
\usepackage[colorlinks]{hyperref}
\hypersetup{%
	plainpages=true,
	breaklinks=true,  
	hypertexnames=false,  
	pageanchor=true,
	colorlinks=true,
	linkcolor={blue},
	citecolor={red},
	urlcolor={blue},
	anchorcolor={black}
}

\begin{document}

\title{Suppressed Energy Relaxation in the Quantum Rabi Model at the Critical Point}
\author{Ye-Hong Chen}
\affiliation{Fujian Key Laboratory of Quantum Information and Quantum Optics, Fuzhou University, Fuzhou 350116, China}%
\affiliation{Department of Physics, Fuzhou University, Fuzhou 350116, China}%
\affiliation{Theoretical Quantum Physics Laboratory, Cluster for Pioneering Research, RIKEN, Wako-shi, Saitama 351-0198, Japan}

\author{Zhi-Cheng Shi}
\affiliation{Fujian Key Laboratory of Quantum Information and Quantum Optics, Fuzhou University, Fuzhou 350116, China}%
\affiliation{Department of Physics, Fuzhou University, Fuzhou 350116, China}

\author{Yu-Ran Zhang}
\affiliation{School of Physics and Optoelectronics, South China University of Technology, Guangzhou 510640, China}

\author{Franco Nori}
\affiliation{Theoretical Quantum Physics Laboratory, Cluster for Pioneering Research, RIKEN, Wako-shi, Saitama 351-0198, Japan}%
\affiliation{Quantum Information Physics Theory Research Team, Center for Quantum Computing, RIKEN, Wako-shi, Saitama 351-0198, Japan}%
\affiliation{Department of Physics, University of Michigan, Ann Arbor, Michigan 48109-1040, USA}

\author{Yan Xia}\thanks{xia-208@163.com}
\affiliation{Fujian Key Laboratory of Quantum Information and Quantum Optics, Fuzhou University, Fuzhou 350116, China}%
\affiliation{Department of Physics, Fuzhou University, Fuzhou 350116, China}

\date{\today}

\begin{abstract}
We derive a modified master equation for the quantum Rabi model in the parameter regime where quantum 
criticality can occur. The modified master equation can avoid some unphysical predictions, such as
excitations in the system at zero temperature and emission of ground-state photons. Due 
to spectrum collapse, we find that there is mostly no energy relaxation in the system at the critical point. 
For the same reason, phase coherence rapidly reduces and vanishes at the critical point.
We analyze the quantum metrological limits of the system in the presence of dephasing.
These results show a strong limitation on the precision of phase-shift estimation.

\end{abstract}


\maketitle

\textit{Introduction.}---The interaction between light and matter is one of the
most fundamental and ubiquitous physical processes.
In order to obtain a complete quantum description of the interaction of matter 
and light, the quantum Rabi model (QRM) has been introduced and studied since the 1930s \cite{Scully1997Book,Agarwal2012Book,Auffeves2013Book}.
It fully describes the dipolar
interaction between a classical monochromatic field and a two-level system, involving both rotating and counter-rotating terms \cite{FriskKockum2019,FornDiza2019,FornDaz2016Np,Yoshihara2016NP,Yoshihara2017PRA,Bosman2017Njpqi,Braumller2017NC,Lv2018PRX,DiStefano2019Np,Cai2021NC,Chen2021NC,Zheng2023PRL}. 
In the presence of counter-rotating terms, this model predicts many
fascinating quantum phenomena, such as the asymmetry of
the vacuum Rabi splitting \cite{Cao2011Njp}, nonclassical photon statistics
\cite{Ashhab2010Pra,Ashhab2013,Chen2021,Chen2024Prl}, and superradiance transition \cite{Ashhab2013,Ridolfo2012,Ridolfo2013,Hwang2015Prl,Shammah2017,Shapira2018,Chen2024CP}.
Note that the occurrence of superradiance transition exhibited by the QRM is accompanied by
a sudden change of photon number \cite{Ashhab2010Pra,Ashhab2013,Hwang2015Prl}.
Such a critical phenomenon has been explored for quantum sensing
because it is sensitive to minute changes of physical
parameters \cite{Wang2014Njp,Chu2021Prl,Garbe2020Prl,Xu2020Sa,Cai2021NC,Chen2021NC,Zheng2023PRL,Hotter2024Prl,Wan2024Prb,Zhu2024Pra}.

This critical phenomenon occurs 
in the dispersive and deep-strong coupling regime \cite{Hwang2015Prl}, i.e., 
the coupling strength and the frequency of the qubit are much larger than the frequency of the harmonic oscillator \cite{FornDiza2019,FriskKockum2019}.
The physical mechanism of energy relaxation and dephasing of the system is complicated in such a 
regime \cite{Shammah2018Pra,FornDiza2019,FriskKockum2019,DeLiberato2009Pra,Beaudoin2011Pra,Settineri2018Pra,Mercurio2023Prl}.
In the standard quantum-optical master equation \cite{Breuer2002Book,Scully1997Book}, the interaction between different components of 
a hybrid quantum system is usually neglected by assuming weak light-matter couplings.
However, this approach can result in unphysical predictions (e,g., excitations in the 
system even at zero temperature \cite{Beaudoin2011Pra,Settineri2018Pra})
when the light-matter interaction enters the regime where the rotating-wave 
approximation is not applicable. 
The standard input-output theory \cite{Ridolfo2012,Garziano2013Pra,Stassi2013Prl,Chen2024CP} also leads to an unphysical prediction: 
the cavity-qubit system can continuously emit photons even in its ground state.
Efforts have been made to solve these problems by modifying the master equation \cite{DeLiberato2009Pra,Beaudoin2011Pra,Rossatto2017Pra,Settineri2018Pra,Shammah2018Pra,DeBernardis2023Pra}.
For instance, in Ref.~\cite{Beaudoin2011Pra}, decoherence is determined by the bath noise spectrum evaluated at the dressed transition
frequencies of the QRM. This approach can correctly describe the relaxation to the thermal equilibrium
density matrix for the system with
anharmonicity larger than the transition linewidths, but can hardly 
be applied to the QRM in the dispersive coupling regime \cite{Settineri2018Pra}.
Therefore, it is an open problem to describe dissipation of the QRM at the critical point because the situation becomes even more complicated due to the 
eigenenergy spectrum collapse. 

Following the idea presented in Ref.~\cite{Settineri2018Pra}, 
we analyze the QRM and derive
the master equation by applying the second-order
Born approximation with the assumption of weak system-bath interaction under the case of an Ohmic bath.
The system operators in terms
of the dressed states of the QRM are decomposed without performing the usual secular
approximation. The results show that 
the energy relaxation decreases exponentially when the system approaches the critical point.
However, phase coherence rapidly decreases and vanishes at the phase transition.
That is, pure dephasing dominates the dissipative dynamics of the system at the critical point.
As a result, the precision of phase estimation is seriously limited.

\textit{Criticality of the QRM.}---We consider a qubit with transition frequency $\Omega_{q}$ coupled with a cavity field with resonant frequency $\omega_{c}$. The system can be described by the QRM,
\begin{align}
	\hat{H}_{R}=\omega_{c}\hat{a}^{\dag}\hat{a}+\frac{\Omega_{q}}{2}\hat{\sigma}_{z}-\lambda(\hat{a}^{\dag}+\hat{a})\hat{\sigma}_{x},
\end{align}
where $\hat{\sigma}_{x,z}$ are the Pauli matrices, $\hat{a}$ ($\hat{a}^{\dag}$) 
is the annihilation (creation) operator for a cavity field, and $\lambda$ is the coupling strength.
In the thermodynamic limit $\Omega_{q}/\omega_{c}\rightarrow \infty$, $\hat{H}_{R}$ can be 
approximatively diagonalized using a Schrieffer-Wolff transformation as \cite{Hwang2015Prl}
\begin{align}
	\hat{H}_{N}=\omega_{c}\hat{a}^{\dag}\hat{a}+\frac{\Omega_{q}}{2}\hat{\sigma}_{z}-\frac{\lambda^2}{\Omega_{q}}(\hat{a}^{\dag}+\hat{a})^{2}\hat{\sigma}_{z},
\end{align}
which
provides a faithful description of the system ground state
in the normal phase (NP). It is stable for $g=2\lambda/\sqrt{\Omega_{q}\omega_{c}}<1$.
By projecting the system onto the ground state $|\!\downarrow\rangle$ of $\hat{\sigma}_{z}$,
we can obtain the low-energy eigenstates of $\hat{H}_{R}$ $|E_{n}\rangle=\hat{S}(r_N)|n\rangle|\!\downarrow\rangle$ with
transition frequency between the eigenstates $|E_{n}\rangle$ and $|E_{m\neq n}\rangle$:
\begin{align}\label{eq3}
	\omega=(m-n)\omega_{c}e^{-2r_{N}}, \ \ \  r_{N}=-\frac{1}{4}\ln(1-g^2).
\end{align}
Here, $\hat{S}(*)=\exp[*/2(\hat{a}^{\dag 2}-\hat{a}^{2})]$ is the squeezing operator.

For $g>1$, the system experiences a phase transition
towards the superradiant phase (SP) \cite{Hwang2015Prl,Chen2024CP}, after transforming the Hamiltonian $H_{R}$ by displacing the cavity field with $\hat{D}(\pm\alpha)=\exp[\pm\alpha(\hat{a}^{\dag}-\hat{a})]$ and $\alpha=\sqrt{(\lambda/\omega_{c})^{2}(1-g^{-4})}$, $\hat{H}_{R}$ can be diagonalized using the same procedure \cite{Hwang2015Prl}, 
\begin{align}
	\hat{H}_{S}=\omega_{c}\hat{a}^{\dag}\hat{a}+\frac{\tilde{\Omega}_{q}}{2}\hat{\tilde{\sigma}}_{z}^{\pm}-\frac{\omega_{c}}{2g^{3}}(\hat{a}^{\dag}+\hat{a})^{2}\hat{\tilde{\sigma}}_{z}^{\pm}+\omega_{c}\alpha^2.
\end{align}
Here, $\tilde{\Omega}_{q}=g^{2}\Omega_{q}$ is the rescaled frequency and 
$\hat{\tilde{\sigma}}_{z}^{\pm}=\left(1/2g^2 \right)\hat{\sigma}_{z}\pm(2\lambda\alpha/\tilde{\Omega}_{q})\hat{\sigma}_{x}$ are the rescaled Pauli matrices. 
Low-energy degenerate eigenstates in the SP are $|E_{n}^{\pm}\rangle=\hat{D}(\pm\alpha)|\hat{S}(r_{S})|n\rangle|\!\!\downarrow^{\pm}\rangle$,
with transition frequency
between $|E_{n}\rangle_{\pm}$ and $|E_{m}\rangle_{\pm}$:
\begin{align}
	\omega=(m-n)\omega_{c}e^{-2r_{S}},\ \ \ r_{S}=-\frac{1}{4}\ln(1-g^{-4}).
\end{align}
Here, $|\!\!\downarrow^{\pm}\rangle$ are the 
ground states of $\hat{\tilde{\sigma}}_{z}^{\pm}$, respectively.

A fascinating quantum phenomenon from the above is that the mean photon number in the ground eigenstate $|E_{0}\rangle$
suddenly reaches infinite, i.e., $\langle E_{0}|\hat{a}^{\dag}\hat{a}|E_{0}\rangle\rightarrow \infty$ 
at the critical point $g=1$. This is also a signature that indicates the occurrence of the quantum 
phase transition \cite{Hwang2015Prl,Chen2024CP}. However, directly detecting such photons 
in experiments are difficult because these photons are virtual \cite{FriskKockum2019,FornDiza2019,Chen2024CP}.
We may assume that the cavity decay can be described by the standard Lindblad form of the master equation
at zero temperature as
\begin{align}\label{eq6}
	\dot{\hat{\rho}}=&-i[\hat{H}_{R},\hat{\rho}]+\mathcal{L}\hat{\rho},
\end{align} 
where $\mathcal{L}\hat{\rho}$ is the Lindblad superoperator. Generally speaking,
it has the standard Lindblad form: $\mathcal{L}\hat{\rho}=\frac{\kappa}{2}\left(2\hat{a}\hat{\rho} \hat{a}^{\dag}-\hat{\rho} \hat{a}^{\dag}\hat{a}-\hat{a}^{\dag}\hat{a}\hat{\rho}\right)$.
However, due to $\hat{a}|E_{0}\rangle\rightarrow \sum_{m}c_{m}|E_{m}\rangle $ in the ultrastrong coupling regime, for the system in the ground state $|E_{0}\rangle$, this master equation leads to unphysical predictions, e.g., production of excitations in the system even at zero
temperature \cite{DeLiberato2009Pra,Beaudoin2011Pra,Rossatto2017Pra,Settineri2018Pra,DeBernardis2023Pra}. Therefore, the master equation, especially, the superoperator $\mathcal{L}\hat{\rho}$,
needs to be modified in such a way that it damps any initial
state toward the actual ground state $|E_{0}\rangle$.

\textit{Generalized master equation with energy relaxation.}---The system
operator in the interaction picture can be described using dressed states by
\begin{align}
	\hat{\tilde{L}}_{k}=\sum_{n,m}\hat{P}_{n}\left(\hat{o}_k+\hat{o}_k^{\dag}\right)\hat{P}_{m}e^{-i\omega t}=\sum_{\omega}\hat{L}_{k}(\omega)e^{-i\omega t},
\end{align}
where  $\hat{P}_{n}=|E_n\rangle\langle E_n|$ is the projector onto the respective eigenspace, $\hat{o}_{k}$ ($\hat{o}_{k}^{\dag}$) are the lowering (rising) operators of the subsystems, and $\omega$ is the transition frequency between the eigenstates $|E_{m}\rangle$ and $|E_{n}\rangle$. For the QRM,
$\hat{o}_{1}=\hat{a}$ and $\hat{o}_{2}=\hat{\sigma}_{-}$.
In this way, the system operators $\hat{o}_{k}$ are expressed as a sum over
transition operators, which cause transitions between energy eigenstates of a hybrid quantum system.
In the interaction picture,
the system-bath interaction Hamiltonian can be written as
$\hat{H}_{\rm SB}=\sum_{k}\hat{\tilde{L}}_{k}\left(\sum_{l}\alpha_{k,l}\hat{b}^{k,l}e^{-i\nu_{l}t}+{\rm h.c.}\right)$, 
where $\hat{b}$ is the bosonic annihilation operator
for the $l$th bath mode with frequency $\nu_{l}$ of the $k$th reservoir and
$\alpha_{k,l}$ is the system-bath coupling strength \cite{Settineri2018Pra}.

Following the standard procedure for a system coupled to 
zero-temperature reservoirs \cite{Scully1997Book,Agarwal2012Book,Breuer2002Book}, we can obtain a Liouvillian superoperator $\mathcal{L}$:%
\begin{small}\begin{align}\label{eq7}
	\mathcal{L}_{\rm g}\hat{\rho}=&\sum_{k,\omega,\omega'}\frac{\Gamma_{k}(\omega)}{2}\left[\hat{L}_{k}(\omega)\hat{\rho} \hat{L}_{k}(-\omega')-\hat{L}_{k}(-\omega')\hat{L}_{k}(\omega)\hat{\rho}\right]\cr 
	 &+\frac{\Gamma_{k}(\omega')}{2}\left[\hat{L}_{k}(\omega)\hat{\rho} \hat{L}_{k}(-\omega')-\hat{\rho} \hat{L}_{k}(-\omega')\hat{L}_{k}(\omega)\right],
\end{align}
\end{small}%
where $\Gamma_{k}(\omega)=2\pi g_{k}(\omega)|\alpha_{k}(\omega)|^2$, with $g_{k}(\omega)$ being the reservoir density of states and $\alpha_{k}(\omega)$ being the system-reservoir coupling strength. 
Note that $\omega$ and $\omega'$ can be positive, negative, or zero. For $\omega>0$, 
$\hat{L}_{k}(\omega)$ takes the system from an eigenstate with higher energy to
one with lower energy and it does the opposite for $\omega<0$. Thus we impose $\omega\geq0$, resulting in
$\hat{L}_{k}(-\omega)=\left[\hat{L}_{k}(\omega)\right]^{\dag}$. 

Zero-frequency
transitions with $\omega=\omega'=0$ give rise to additional pure dephasing contributions
that can be regarded as a generalization of those
appearing in the master equation for optomechanical systems
in the ultrastrong-coupling regime \cite{Hu2015Pra}.
This is very important for describing the system dissipative dynamics
at the critical point of the QRM due to the spectrum collapse at this point.
That is, at the critical point, there is no energy relaxation in the system but only additional pure dephasing with a rate
\begin{align}\label{eq10}
	\Gamma_{\phi}^{\pm}=\Gamma_{k}(0)=\int_{0}^{t}d\tau\int_{0}^{\infty}d\nu g_{k}(\nu)|\alpha_{k}(\nu)|^{2}e^{\pm i\nu \tau}.
\end{align}
Here, for the particular case of an Ohmic
bath, $g_{k}(\nu)|\alpha_{k}(\nu)|^{2}=\gamma_{k}\nu /(2\pi f_{k})$, with $\gamma_{k}$ and $f_{k}$
being, respectively, the energy relaxation and the
frequency of the considered subsystem. In this case, we obtain $\Gamma_{\phi}^{\pm}=0$. 
This interesting result implies that \textit{there is no energy relaxation at the critical point.}

\begin{figure}
	\centering
	\scalebox{0.33}{\includegraphics{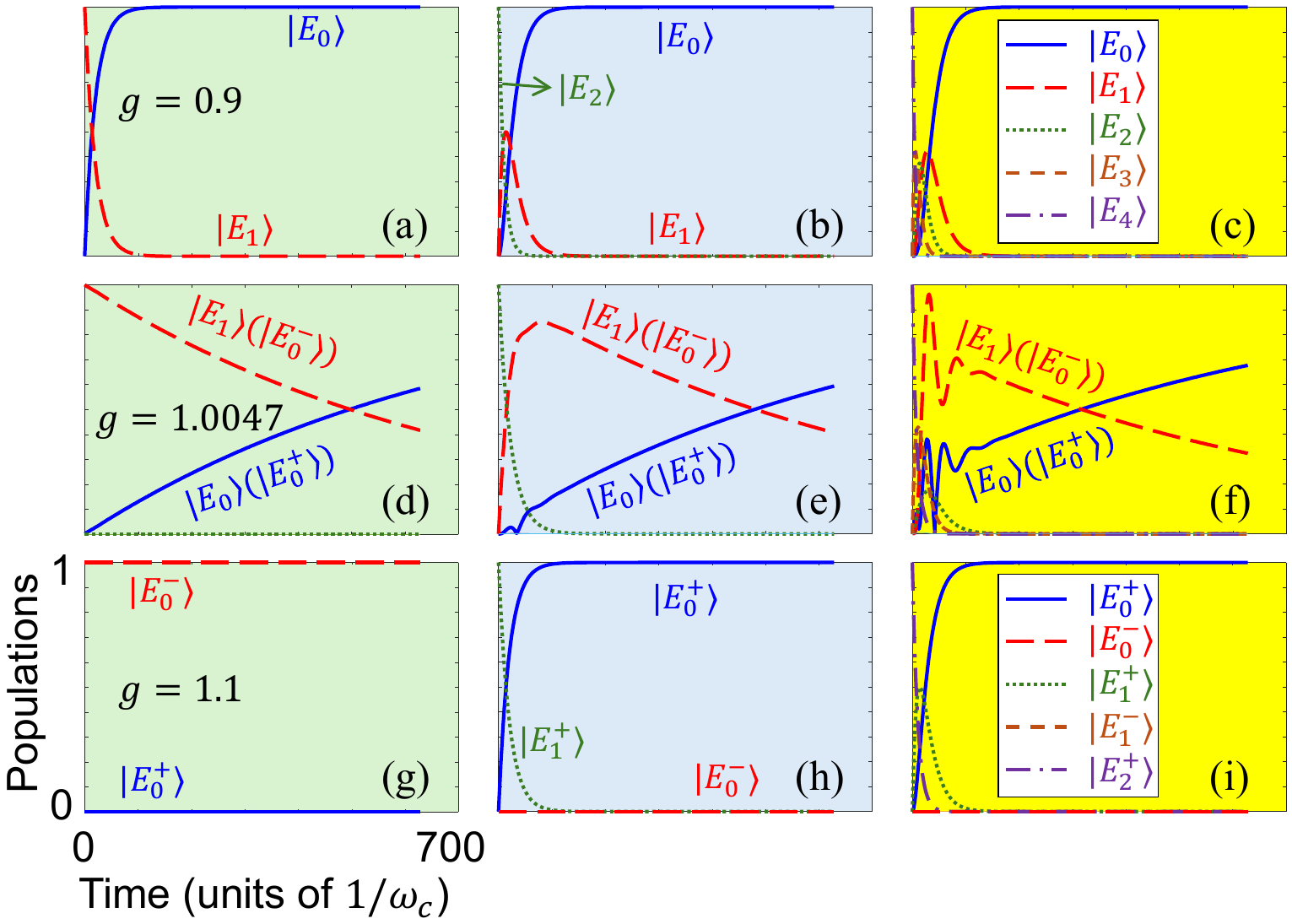}}
	\caption{Populations of the lowest five eigenstates in the presence of cavity decay for different initial states and coupling strengths. The dissipative dynamics of the QRM is governed by Eq.~(\ref{eq6}) with the modified superoperator $\mathcal{L}_{\rm g}\rho$ in Eq.~(\ref{eq7}). The green-, blue-, and yellow-shaded plots denote the dynamics with the initial states $|E_{1}\rangle$($|E_{0}^{-}\rangle$), $|E_{2}\rangle$($|E_{1}^{+}\rangle$), and $|E_{4}\rangle$($|E_{2}^{+}\rangle$), respectively.
		(a--c) Normal phase with $g=0.9$.
	(d--e) Near the critical point with $g=1.0047$.
	(g--i) Superradiant phase with $g=1.1$.	
    We choose $\Omega_{q}/\omega_{c}=10^{4}$ to approximatively reach the thermodynamic limit. The cavity decay parameter is $\gamma_{1}=0.05\omega_{c}$ and the qubit decay parameter is $\gamma_{2}=0.01\omega_{c}$. 
    Note that the critical point is approximatively $g\simeq 1.0047$ for finite frequencies.}
	\label{fig1}
\end{figure}

However, because $\Omega_{q}/\omega_{c}$ is finite, it is in fact
impossible for the system to accurately reach the critical point with $\omega=\omega'=0$.
Also considering the case of an Ohmic
bath, we can calculate the dissipation rate for $\Gamma_{1}(\omega)=\gamma_{1}\omega/\omega_{c}$, corresponding to the cavity decay, and 
the dissipation rate $\Gamma_{2}(\omega)=\gamma_{2}\omega/\Omega_{q}$, for the qubit decay. 
In the $\Omega_{q}/\omega_{c}\rightarrow \infty$ limit,
it is obvious that $\omega/\Omega_{q}\rightarrow 0$, resulting in $\Gamma_{2}(\omega)\rightarrow0$.
Therefore, the influence of qubit decay becomes negligible.
We can thus focus on analyzing the influence of the cavity decay.

In the NP with $g<1$, we have 
$\Gamma_{1}(\omega)=\gamma_{1}(m-n)\sqrt{1-g^{2}}$, according to Eq.~(\ref{eq3}),
which decreases when increasing $g$. 
The dressed cavity operator $\hat{L}_{1}(\omega)$ is
\begin{align*}
	\hat{L}_{1}(\omega)=|E_{n}\rangle\langle E_{m}|\langle n|\hat{S}^{\dag}(r_{N})(\hat{a}+\hat{a}^{\dag})\hat{S}(r_{N})|m\rangle.
\end{align*}
Then, assuming $\omega_{c}\gg\gamma_{1}$ and
neglecting the fast oscillating terms in Eq.~(\ref{eq7}) 
by the rotating-wave approximation, Eq.~(\ref{eq7}) becomes
\begin{align}\label{eq11}
	\mathcal{L}_{\rm g}\hat{\rho}&\approx \sum_{n}\gamma_{1}(1-g^{2})(n+1)\mathcal{D}[| E_{n}\rangle\langle E_{n+1}|]\hat{\rho},
\end{align}
where $\mathcal{D}[\hat{o}]\hat{\rho}=\hat{o}\hat{\rho} \hat{o}^{\dag}-\left(\hat{o}^{\dag}\hat{o}\hat{\rho}+\hat{\rho} \hat{o}^{\dag}\hat{o}\right)/2$ is the Lindblad
superoperator. 
In Fig.~\ref{fig1}, according to Eq.~(\ref{eq7}), we show the dissipative evolutions of the system for different coupling strengths when the system is initially in different eigenstates. 

In the NP, the energy levels are well separated. The dissipator
in Eq.~(\ref{eq7}) can be regarded to be in Lindblad form, which describes 
the relaxation from a higher-energy state to a lower-energy state [see Figs.~\ref{fig1}(a--c)]. 
When the normalized coupling strength $g$ approaches the critical point, i.e., $g\simeq 1.0047$,
we can see from Figs.~\ref{fig1}(d-e) that it takes a much longer time for the system to become stable, i.e., the decay rate becomes smaller.
The critical point slightly shifts to $g\simeq 1.0047$ because the frequencies are finite and the thermodynamic limit cannot be fully satisfied.
Moreover, the decay cannot vanish as predicted in Eq.~(\ref{eq11}) also because of the finite frequencies.

In the SP with $g>1$, the cavity decay rate becomes 
$\Gamma_{1}(\omega)=\gamma_{1}(m-n)\sqrt{1-g^{-4}}$, which increases when increasing $g$.
The dressed cavity operator becomes $\hat{L}_{1}^{l,l'}(\omega)=\tilde{\Gamma}_{n,m}^{l,l'}|E_{n}^{l}\rangle\langle E_{m}^{l'}|$, where $l,l'=\pm$, and
\begin{align*}
	\tilde{\Gamma}_{n,m}^{l,l'}=\langle n|\hat{S}^{\dag}(r_{S})\hat{D}^{\dag}(l\alpha)(\hat{a}+\hat{a}^{\dag})\hat{D}(l'\alpha)\hat{S}(r_{S})|m\rangle.
\end{align*}
For large $\alpha$, $\tilde{\Gamma}_{n,m}^{l,l'}$ has significant values when $l=l'$. 
The transition obeys a parity conservation that when the system is initially in a state with $l=+$, 
it can hardly decay to a lower-energy state with $l=-$ [see Figs.~\ref{fig1}(g--h)]. 
For instance, when the system is initially in $|E_{0}^{-}\rangle$, it remains in the initial state in the presence of 
cavity decay [see Fig.~\ref{fig1}(g)].
Hence, the dressed cavity operator can be further simplified as
$\hat{L}_{1}^{\pm}(\omega)=|E_{n}^{\pm}\rangle\langle E_{n+1}^{\pm}|\exp(-r_{S})\sqrt{n+1}$.
Therefore, Eq.~(\ref{eq7}) becomes
\begin{align}\label{eq12}
	\mathcal{L}_{\rm g}\hat{\rho}\approx\sum_{l}\sum_{n}\gamma_{1}(1-g^{-4})(n+1)\mathcal{D}[|E_{n}^{l}\rangle\langle E_{n+1}^{l}|]\hat{\rho}.
\end{align}
The result obtained from Eqs.~(\ref{eq11},\ref{eq12}) coincides with the analysis below Eq.~(\ref{eq10}): energy relaxation vanishes at the critical point with $g=1$. 
However, due to the finite frequencies, the critical point with $g=1$ can never be reached.
It is only possible to observe a significant decrease of decay rate as shown in Figs.~\ref{fig1}(d--f).

\begin{figure}
	\centering
	\scalebox{0.32}{\includegraphics{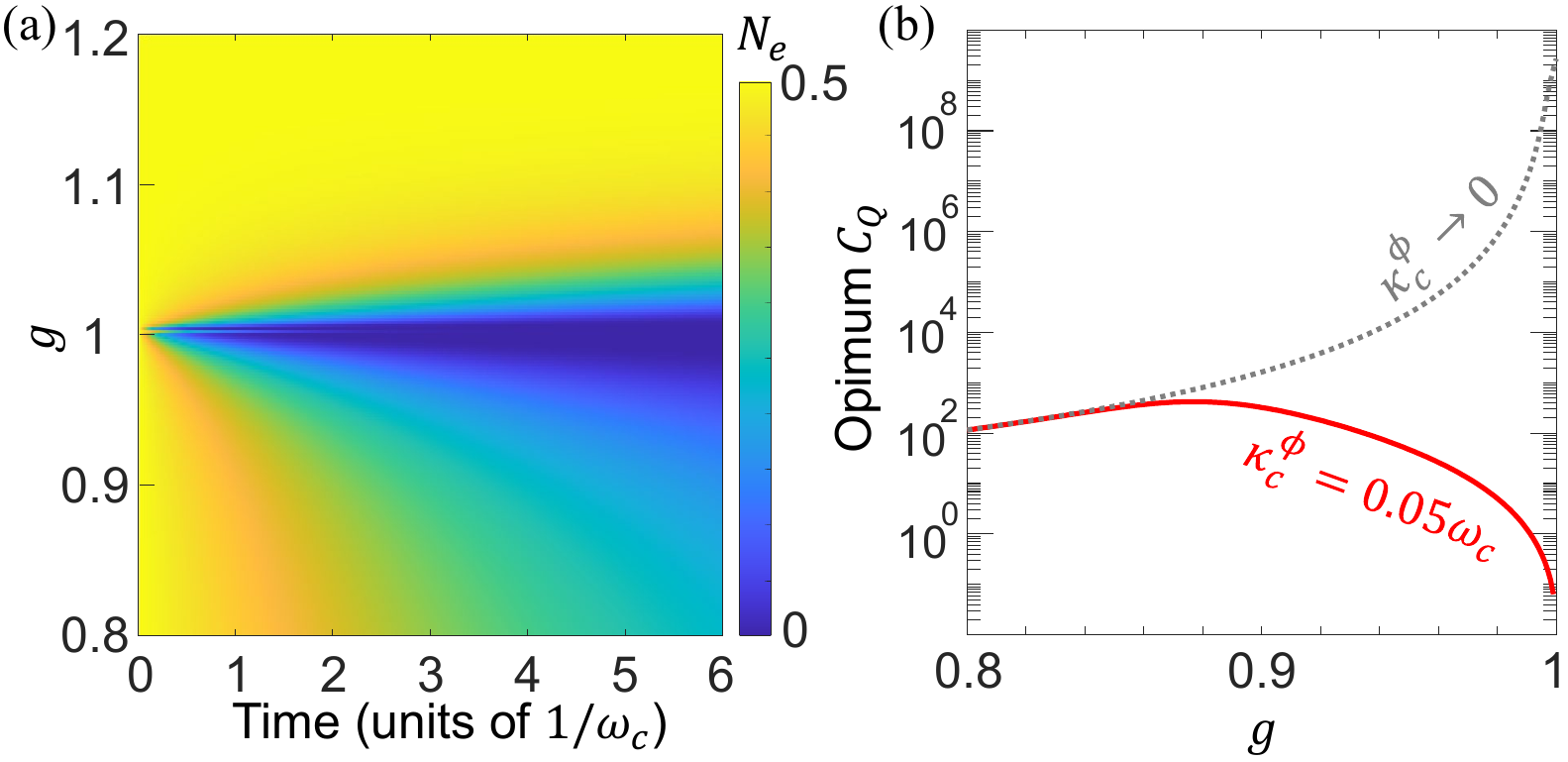}}
	\caption{(a) Time evolution of the nondiagonal element $N_{e}=|\langle E_{2}|\hat{\rho}(t)|E_{0}\rangle|$ versus the normalized coupling strength $g$, where the density matrix $\hat{\rho}(t)$ is calculated based on the master equation in Eq.~(\ref{eq6}) with the superoperator $\mathcal{L}_{\phi}\hat{\rho}$ in Eq.~(\ref{eq13}).
		The dephasing parameters are $\kappa_{c}^{\phi}=\kappa_{q}^{\phi}=0.05\omega_{c}$. (b) Upper bound of the QFI calculated for different $\kappa_{c}^{\phi}$. The evolution state $|\psi_{S,E}(\Phi)\rangle$ is obtained using $H_{R}$ with initial state $|\psi_{S}\rangle=(|E_{0}\rangle+|E_{2}\rangle)/\sqrt{2}$ and the accumulated 
		phase shift is assumed to be $\Phi=\pi/4$. We choose $\Omega_{q}/\omega_{c}=10^{4}$ to approximatively reach the thermodynamic limit.}
	\label{fig2}
\end{figure}

\textit{Pure dephasing.}---In the NP, the finite-frequency effect allows to consider
stochastic functions with a low-frequency spectral density, i.e., the eigenstates can be separated well.
The superoperator $\mathcal{L}_{\phi}\hat{\rho}$ describing pure dephasing becomes
\begin{align}\label{eq13}
	\mathcal{L}_{\phi}\hat{\rho}=&\frac{\kappa^{\phi}_{c}}{2}\mathcal{D}\left[\sum_{n}|E_{n}\rangle\langle E_{n}|\langle E_{n}|\hat{a}^{\dag}\hat{a}|E_{n}\rangle\right]\hat{\rho}\cr
	                       &+\frac{\kappa^{\phi}_{q}}{2}\mathcal{D}\left[\sum_{n}|E_{n}\rangle\langle E_{n}|\langle E_{n}|\hat{\sigma}_{z}|E_{n}\rangle\right]\hat{\rho},
\end{align}
where $\kappa_{c,(q)}^{\phi}$ are dephasing rates. 
For low-energy eigenstates, we have 
$\langle E_{n}|\hat{a}^{\dag}\hat{a}|E_{n}\rangle=n\cosh(2r_{N})+\sinh^{2}(r_{N})$ and $\langle E_{n}|\hat{\sigma}_{z}|E_{n}\rangle=-1$. 
Therefore, when the system approaches the critical point, the sudden increase of $r_{N}$ leads to
a rapid increase of the influence of pure dephasing.
Equation~(\ref{eq13}) is thus simplified as 
\begin{align}\label{eq14}
	\mathcal{L}_{\phi}\hat{\rho}= \frac{\kappa_{c}^{\phi}}{2}\cosh^{2}(2r_{N})\mathcal{D}\left[\sum_{n}n|E_{n}\rangle\langle E_{n}|\right]\hat{\rho}.
\end{align}
The dephasing rate is nearly infinite when the system approaches the critical point.
Note that pure dephasing mainly affects the coherence of the system, which is described by the nondiagonal elements of the density matrix, e.g., $\langle E_{n}|\hat{\rho}(t)|E_{m\neq n}\rangle$.
We now assume the initial state to be $(|E_{0}\rangle+|E_{2}\rangle)/\sqrt{2}$ and show the nondiagonal element $N_{e}=|\langle E_{2}|\hat{\rho}(t)|E_{0}\rangle|$ of the density matrix in Fig.~\ref{fig2}(a).
When $g\rightarrow 1$, the value of the nondiagonal element $N_{e}$ rapidly decreases, 
indicating that the coherence time of the system decreases rapidly near the critical point.

For the SP, we obtain a similar result because $\langle E_{n}^{l}|\hat{a}^{\dag}\hat{a}|E_{n}^{l}\rangle=n\cosh(2r_{S})+\sinh^{2}(r_{S})+|\alpha|^2$ and
$\langle E_{n}^{+}|\hat{\sigma}_{z}|E_{n}^{+}\rangle=\langle E_{n}^{-}|\hat{\sigma}_{z}|E_{n}^{-}\rangle={\rm constant}$.
Pure dephasing in the SP is described by
\begin{align}
	\mathcal{L}_{\phi}\hat{\rho}= \frac{\kappa_{c}^{\phi}}{2}\cosh^{2}(2r_{S})\mathcal{D}\left[\sum_{l}\sum_{n}n|E_{n}^{l}\rangle\langle E_{n}^{l}|\right]\hat{\rho},
\end{align}
which also indicates that pure dephasing approaches infinity at the critical point [see Fig.~\ref{fig2}(a)].

\textit{Quantum metrological limits.}---Phase estimation is a central problem in quantum 
metrology \cite{Giovannetti2011Nat,Genoni2011Prl,Zhang2014Pra,Liu2015Nc}. Dephasing is
a relevant source of noise in optical phase measurements
and must be taken into account. 
We now assume that the system is initially in a pure state $\hat{\rho}(0)=\hat{\rho}_{S}(0)=|\psi_{S}\rangle\langle\psi_{S}|$, where $|\psi_{S}\rangle=\sum_{n}C_{n}|E_{n}\rangle$, with arbitrary constants $C_{n}$ satisfying $\sum_{n}|C_{n}|^{2}=1$.
In the presence of dephasing, the resulting state of the probe in the Markov limit is given by
\begin{align}
	\hat{\rho}_{S}(\Phi)=\sum_{m,n=0}^{\infty}\rho_{m,n}e^{-i\Phi(m-n)-\beta^{2}(m-n)^2}|E_{m}\rangle\langle E_{n}|,
\end{align}
where  $\rho_{m,n}$ is the matrix element of the initial state, $\beta\propto \cosh(2 r_{N})\sqrt{\kappa_{c}^{\phi}/2}$ quantifies the degree of dephasing, and $\Phi$ is the accumulated
phase shift through a coherent evolution.
For any unbiased estimation
the statistical uncertainty is limited by the Cram\'{e}r-Rao bound \cite{Cramer1946book,Rao1973book,Holevo2011Book}.
Assuming that the experiment has $\nu$ repetitions, the quantum version of this bound is
$\delta \Phi\geq 1/\sqrt{\nu F_{Q}(\Phi)}$, where $F_{Q}(\Phi)$ is the quantum Fisher information (QFI) \cite{Ma2011PR,Degen2017Rmp,Lu2021Prl,Xu2022Prl,Zhang2023Pra}.
When $\rho_{S}$ is not a pure state, to apply the symmetric logarithmic derivative approach for calculating the QFI \cite{Note_eq16}, we need to enlarge the original Hilbert space and construct a pure state $\hat{\rho}_{S,E}(\Phi)=|\psi_{S,E}\rangle\langle \psi_{S,E}|$,
where $S$ and $E$ stand for ``system'' and ``environment'', respectively.

Noting that taking the trace over the environment discards information on the full space,
one can obtain a physically motivated upper bound of the QFI as \cite{Escher2012Prl}
\begin{align}
	C_{Q}[\hat{\rho}_{S,E}(\Phi)]\equiv F_{Q}[\hat{\rho}_{S,E}(\Phi)]\geq F_{Q}[\hat{\rho}_{S}(\Phi)].
\end{align}
Therefore, the best upper bound can be calculated by 
the minimum of $C_{Q}$ over all possible purifications of
$\rho_{S}$. To do so, we define a unitary operator $\hat{U}_{E}(\Phi)$ and a Hermitian operator $\hat{h}_{E}(\Phi)=i[\partial_{\Phi} \hat{U}_{E}^{\dag}(\Phi)]\hat{U}_{E}(\Phi)$ 
acting only on the subspace of environment. Moreover, we define another Hermitian operator $\hat{H}_{S,E}(\Phi)$
acts on the full space and satisfies $i\partial_{\Phi}|\psi_{S,E}\rangle=\hat{H}_{S,E}(\Phi)|\psi_{S,E}\rangle$. 
The upper bound is given as $C_{Q}=4\langle[\hat{\mathcal{H}}-\langle \hat{\mathcal{H}}\rangle_{\psi}]^{2}\rangle_{\psi}$, where $\hat{\mathcal{H}}=\hat{H}_{S,E}(\Phi)-\hat{h}_{E}(\Phi)$ and
the averages are
taken over $|\psi_{S,E}\rangle$.
Calculating QFI becomes possible by minimizing $C_{Q}$ over all Hermitian operators $\hat{h}_{E}(\Phi)$ \cite{note_eq17,Boyd2009book}.

Note that $\hat{H}_{R}$ in the NP can be
regarded as a generic harmonic oscillator, i.e., $\hat{H}_{R}\approx\omega_{c}\sqrt{1-g^{2}}{\hat{c}^{\dag}\hat{c}\otimes|\downarrow\rangle\langle \downarrow|}$,
where $\hat{c}=\cosh(r_{N})\hat{a}+\sinh(r_{N})\hat{a}^{\dag}$.
The superoperator $D[*]\hat{\rho}$ describing pure dephasing in Eq.~(\ref{eq14}) can be transformed as $\mathcal{D}[\hat{c}^{\dag}\hat{c}]\hat{\rho}$.
Therefore, we can model the dephasing of the initial probe state through the effect of
the radiation pressure on one of the interferometer mirrors.
The combined-system state is 
\begin{align*}
	|\psi_{S,E}(\Phi)\rangle=\exp{({-i\Phi \hat{c}^{\dag}\hat{c}})}\exp{(i2\beta \hat{c}^{\dag}\hat{c}\hat{x}_{E})}|\psi_{S}\rangle|0_{E}\rangle,
\end{align*}
where $\hat{x}_{E}$ is the dimensionless position
operator and $|0_{E}\rangle$ is the initial state of the environment.
The accumulated phase is then $\Phi=\omega_{c}\sqrt{1-g^2}t$.
To calculate $C_{Q}$, we first calculate the reduced density matrix
of the environment associated with the purification $|\psi_{S,E}(\Phi)\rangle$:
$\hat{\rho}_{E}=\sum_{n}|\rho_{n,n}|^{2}|i\sqrt{2}\beta n_c\rangle\langle i\sqrt{2}\beta n_c|$,
where $|i\sqrt{2}\beta n_c\rangle$ is a coherent state and $n_c=\langle \hat{c}^{\dag}\hat{c}\rangle$.
The optimum Hermitian operator $\hat{h}_{E}(\Phi)$ needs to satisfy
\begin{align*}
	{\rm Tr}_{S}[\mathcal{M}_{S,E}]=\frac{1}{2}\left[{\hat{h}_{E}(\Phi)\hat{\rho}_{E}(\Phi)+\hat{\rho}_{E}(\Phi)\hat{h}_{E}}(\Phi)\right],\cr
	\mathcal{M}_{S,E}=\frac{i}{2}\left(|\partial_{\Phi}\psi_{S,E}\rangle\langle\psi_{S,E}|-|\psi_{S,E}\rangle\langle\partial_{\Phi}\psi_{S,E}|\right),
\end{align*}
so that $C_{Q}$ is minimum \cite{Escher2012Prl}.
A possible solution for $\hat{h}_{E}$ is $\hat{h}_{E}=i\zeta\hat{p}_{E}/(2\beta)$ when $\sqrt{2}\beta n_c\gg1$, where $\zeta$ is a variational parameter and $\hat{p}_{E}$ is the dimensionless momentum
operator of the environment. The upper bound of the QFI is then $C_{Q}=(1-\zeta^2)4\Delta n_c^{2}+\zeta^2/(2\beta^2)$,
which reaches its minimum value when $\zeta=8\Delta n_c^{2}\beta^2/(1+8\Delta n_c^{2}\beta^2)$. Here,
$\Delta n_c^2=\langle \hat{c}^{\dag}\hat{c}\hat{c}^{\dag}\hat{c}\rangle- \langle \hat{c}^{\dag}\hat{c}\rangle^2$ is the variance of the operator $\hat{c}^{\dag}\hat{c}$ in the
initial probe state \cite{Pang2017Nc,Chu2021Prl,Garbe2020Prl}. At the critical point, we have $\beta\rightarrow \infty$, resulting
in $\zeta\rightarrow 1$ and $C_{Q}\rightarrow 0$ [see Fig.~\ref{fig2}(b)].
This is different from the case of a pure state with $\beta\rightarrow0$, where $C_{Q}$ exponentially increases when $g\rightarrow 1$ [see Fig.~\ref{fig2}(b)] \cite{Chu2021Prl,Garbe2020Prl,Hotter2024Prl}.
A nontrivial bound for the
precision of phase estimation in the presence of dephasing, valid for any input state, is
\begin{align}
	\delta\Phi\geq\sqrt{\frac{1}{\nu C_{Q}}}= \sqrt{\frac{1}{4\nu\Delta n_{c}^2}+\frac{2\beta^2}{\nu}},
\end{align}
which approaches infinity when $g\rightarrow1$.
The exponential increase of the dephasing rate seriously
affects the
precision of the phase estimation when considering dephasing.

\textit{Conclusions.}---We have studied dissipation in the QRM
at the critical point. 
The master equation
has been modified in such a way that it damps any initial
state toward the actual ground state.
According to the modified master equation, we have found the interesting result that 
the rate of energy relaxation decreases when the system approaches the critical point and
\textit{it can even vanish when the system is exactly at the critical point.}
This is because the rate of relaxation between two eigenstates depends
on the noise spectral density at frequency $\omega$, and
the spectrum collapses at the critical point, leading to $\omega=0$.
However, pure dephasing has great influence on the system at the critical point.
The rate of dephasing can even reach infinity when the system parameters reach the thermodynamic limit.
This significantly limits the precision of the phase estimation.
We hope that our results help in determining the fundamental precision limits of
critical quantum metrology based on the QRM.

\begin{acknowledgments}

Y.-H.C. is supported by the National Natural Science Foundation of China
under Grant No. 12304390.
Y.X. is supported
by the National Natural Science Foundation of China
under Grant No. 11575045 and No. 62471143, the Natural Science Funds
for Distinguished Young Scholar of Fujian Province under
Grant 2020J06011 and Project from Fuzhou University
under Grant JG202001-2.
F.N. is supported in part by: 
Nippon Telegraph and Telephone Corporation (NTT) Research, 
the Japan Science and Technology Agency (JST) 
[via the CREST Quantum Frontiers program Grant No. JPMJCR24I2, 
the Quantum Leap Flagship Program (Q-LEAP), and the Moonshot R\&D Grant Number JPMJMS2061], 
and the Office of Naval Research (ONR) Global (via Grant No. N62909-23-1-2074).

\end{acknowledgments}

\bibliography{references}

\begin{thebibliography}{61}%
\makeatletter
\providecommand \@ifxundefined [1]{%
 \@ifx{#1\undefined}
}%
\providecommand \@ifnum [1]{%
 \ifnum #1\expandafter \@firstoftwo
 \else \expandafter \@secondoftwo
 \fi
}%
\providecommand \@ifx [1]{%
 \ifx #1\expandafter \@firstoftwo
 \else \expandafter \@secondoftwo
 \fi
}%
\providecommand \natexlab [1]{#1}%
\providecommand \enquote  [1]{``#1''}%
\providecommand \bibnamefont  [1]{#1}%
\providecommand \bibfnamefont [1]{#1}%
\providecommand \citenamefont [1]{#1}%
\providecommand \href@noop [0]{\@secondoftwo}%
\providecommand \href [0]{\begingroup \@sanitize@url \@href}%
\providecommand \@href[1]{\@@startlink{#1}\@@href}%
\providecommand \@@href[1]{\endgroup#1\@@endlink}%
\providecommand \@sanitize@url [0]{\catcode `\\12\catcode `\$12\catcode
  `\&12\catcode `\#12\catcode `\^12\catcode `\_12\catcode `\%12\relax}%
\providecommand \@@startlink[1]{}%
\providecommand \@@endlink[0]{}%
\providecommand \url  [0]{\begingroup\@sanitize@url \@url }%
\providecommand \@url [1]{\endgroup\@href {#1}{\urlprefix }}%
\providecommand \urlprefix  [0]{URL }%
\providecommand \Eprint [0]{\href }%
\providecommand \doibase [0]{http://dx.doi.org/}%
\providecommand \selectlanguage [0]{\@gobble}%
\providecommand \bibinfo  [0]{\@secondoftwo}%
\providecommand \bibfield  [0]{\@secondoftwo}%
\providecommand \translation [1]{[#1]}%
\providecommand \BibitemOpen [0]{}%
\providecommand \bibitemStop [0]{}%
\providecommand \bibitemNoStop [0]{.\EOS\space}%
\providecommand \EOS [0]{\spacefactor3000\relax}%
\providecommand \BibitemShut  [1]{\csname bibitem#1\endcsname}%
\let\auto@bib@innerbib\@empty
\bibitem [{\citenamefont {Scully}\ and\ \citenamefont
  {Zubairy}(1997)}]{Scully1997Book}%
  \BibitemOpen
  \bibfield  {author} {\bibinfo {author} {\bibfnamefont {M.~O.}\ \bibnamefont
  {Scully}}\ and\ \bibinfo {author} {\bibfnamefont {M.~S.}\ \bibnamefont
  {Zubairy}},\ }\href {\doibase 10.1017/cbo9780511813993} {\emph {\bibinfo
  {title} {Quantum Optics}}}\ (\bibinfo  {publisher} {Cambridge University
  Press},\ \bibinfo {address} {Cambridge, England},\ \bibinfo {year}
  {1997})\BibitemShut {NoStop}%
\bibitem [{\citenamefont {Agarwal}(2012)}]{Agarwal2012Book}%
  \BibitemOpen
  \bibfield  {author} {\bibinfo {author} {\bibfnamefont {G.~S.}\ \bibnamefont
  {Agarwal}},\ }\href {\doibase 10.1017/cbo9781139035170} {\emph {\bibinfo
  {title} {Quantum Optics}}}\ (\bibinfo  {publisher} {Cambridge University
  Press},\ \bibinfo {address} {Cambridge, England},\ \bibinfo {year}
  {2012})\BibitemShut {NoStop}%
\bibitem [{\citenamefont {Auffeves}\ \emph {et~al.}(2013)\citenamefont
  {Auffeves} \emph {et~al.}}]{Auffeves2013Book}%
  \BibitemOpen
  \bibfield  {author} {\bibinfo {author} {\bibfnamefont {A.}~\bibnamefont
  {Auffeves}} \emph {et~al.},\ }\href {https://doi.org/10.1142/8758} {\emph
  {\bibinfo {title} {Strong light-matter coupling: from atoms to solid-state
  systems}}}\ (\bibinfo  {publisher} {World Scientific},\ \bibinfo {year}
  {2013})\BibitemShut {NoStop}%
\bibitem [{\citenamefont {Kockum}\ \emph {et~al.}(2019)\citenamefont {Kockum},
  \citenamefont {Miranowicz}, \citenamefont {Liberato}, \citenamefont
  {Savasta},\ and\ \citenamefont {Nori}}]{FriskKockum2019}%
  \BibitemOpen
  \bibfield  {author} {\bibinfo {author} {\bibfnamefont {A.~F.}\ \bibnamefont
  {Kockum}}, \bibinfo {author} {\bibfnamefont {A.}~\bibnamefont {Miranowicz}},
  \bibinfo {author} {\bibfnamefont {S.~De}\ \bibnamefont {Liberato}}, \bibinfo
  {author} {\bibfnamefont {S.}~\bibnamefont {Savasta}}, \ and\ \bibinfo
  {author} {\bibfnamefont {F.}~\bibnamefont {Nori}},\ }\bibfield  {title}
  {\enquote {\bibinfo {title} {Ultrastrong coupling between light and
  matter},}\ }\href {\doibase 10.1038/s42254-018-0006-2} {\bibfield  {journal}
  {\bibinfo  {journal} {Nat. Rev. Phys.}\ }\textbf {\bibinfo {volume} {1}},\
  \bibinfo {pages} {19--40} (\bibinfo {year} {2019})}\BibitemShut {NoStop}%
\bibitem [{\citenamefont {Forn-D\'{\i}az}\ \emph {et~al.}(2019)\citenamefont
  {Forn-D\'{\i}az}, \citenamefont {Lamata}, \citenamefont {Rico}, \citenamefont
  {Kono},\ and\ \citenamefont {Solano}}]{FornDiza2019}%
  \BibitemOpen
  \bibfield  {author} {\bibinfo {author} {\bibfnamefont {P.}~\bibnamefont
  {Forn-D\'{\i}az}}, \bibinfo {author} {\bibfnamefont {L.}~\bibnamefont
  {Lamata}}, \bibinfo {author} {\bibfnamefont {E.}~\bibnamefont {Rico}},
  \bibinfo {author} {\bibfnamefont {J.}~\bibnamefont {Kono}}, \ and\ \bibinfo
  {author} {\bibfnamefont {E.}~\bibnamefont {Solano}},\ }\bibfield  {title}
  {\enquote {\bibinfo {title} {Ultrastrong coupling regimes of light-matter
  interaction},}\ }\href {\doibase 10.1103/RevModPhys.91.025005} {\bibfield
  {journal} {\bibinfo  {journal} {Rev. Mod. Phys.}\ }\textbf {\bibinfo {volume}
  {91}},\ \bibinfo {pages} {025005} (\bibinfo {year} {2019})}\BibitemShut
  {NoStop}%
\bibitem [{\citenamefont {Forn-Díaz}\ \emph {et~al.}(2016)\citenamefont
  {Forn-Díaz}, \citenamefont {García-Ripoll}, \citenamefont {Peropadre},
  \citenamefont {Orgiazzi}, \citenamefont {Yurtalan}, \citenamefont
  {Belyansky}, \citenamefont {Wilson},\ and\ \citenamefont
  {Lupascu}}]{FornDaz2016Np}%
  \BibitemOpen
  \bibfield  {author} {\bibinfo {author} {\bibfnamefont {P.}~\bibnamefont
  {Forn-Díaz}}, \bibinfo {author} {\bibfnamefont {J.~J.}\ \bibnamefont
  {García-Ripoll}}, \bibinfo {author} {\bibfnamefont {B.}~\bibnamefont
  {Peropadre}}, \bibinfo {author} {\bibfnamefont {J.-L.}\ \bibnamefont
  {Orgiazzi}}, \bibinfo {author} {\bibfnamefont {M.~A.}\ \bibnamefont
  {Yurtalan}}, \bibinfo {author} {\bibfnamefont {R.}~\bibnamefont {Belyansky}},
  \bibinfo {author} {\bibfnamefont {C.~M.}\ \bibnamefont {Wilson}}, \ and\
  \bibinfo {author} {\bibfnamefont {A.}~\bibnamefont {Lupascu}},\ }\bibfield
  {title} {\enquote {\bibinfo {title} {Ultrastrong coupling of a single
  artificial atom to an electromagnetic continuum in the nonperturbative
  regime},}\ }\href {\doibase 10.1038/nphys3905} {\bibfield  {journal}
  {\bibinfo  {journal} {Nat. Phys.}\ }\textbf {\bibinfo {volume} {13}},\
  \bibinfo {pages} {39–43} (\bibinfo {year} {2016})}\BibitemShut {NoStop}%
\bibitem [{\citenamefont {Yoshihara}\ \emph {et~al.}(2016)\citenamefont
  {Yoshihara}, \citenamefont {Fuse}, \citenamefont {Ashhab}, \citenamefont
  {Kakuyanagi}, \citenamefont {Saito},\ and\ \citenamefont
  {Semba}}]{Yoshihara2016NP}%
  \BibitemOpen
  \bibfield  {author} {\bibinfo {author} {\bibfnamefont {F.}~\bibnamefont
  {Yoshihara}}, \bibinfo {author} {\bibfnamefont {T.}~\bibnamefont {Fuse}},
  \bibinfo {author} {\bibfnamefont {S.}~\bibnamefont {Ashhab}}, \bibinfo
  {author} {\bibfnamefont {K.}~\bibnamefont {Kakuyanagi}}, \bibinfo {author}
  {\bibfnamefont {S.}~\bibnamefont {Saito}}, \ and\ \bibinfo {author}
  {\bibfnamefont {K.}~\bibnamefont {Semba}},\ }\bibfield  {title} {\enquote
  {\bibinfo {title} {Superconducting qubit{\textendash}oscillator circuit
  beyond the ultrastrong-coupling regime},}\ }\href {\doibase
  10.1038/nphys3906} {\bibfield  {journal} {\bibinfo  {journal} {Nat. Phys.}\
  }\textbf {\bibinfo {volume} {13}},\ \bibinfo {pages} {44--47} (\bibinfo
  {year} {2016})}\BibitemShut {NoStop}%
\bibitem [{\citenamefont {Yoshihara}\ \emph {et~al.}(2017)\citenamefont
  {Yoshihara}, \citenamefont {Fuse}, \citenamefont {Ashhab}, \citenamefont
  {Kakuyanagi}, \citenamefont {Saito},\ and\ \citenamefont
  {Semba}}]{Yoshihara2017PRA}%
  \BibitemOpen
  \bibfield  {author} {\bibinfo {author} {\bibfnamefont {F.}~\bibnamefont
  {Yoshihara}}, \bibinfo {author} {\bibfnamefont {T.}~\bibnamefont {Fuse}},
  \bibinfo {author} {\bibfnamefont {S.}~\bibnamefont {Ashhab}}, \bibinfo
  {author} {\bibfnamefont {K.}~\bibnamefont {Kakuyanagi}}, \bibinfo {author}
  {\bibfnamefont {S.}~\bibnamefont {Saito}}, \ and\ \bibinfo {author}
  {\bibfnamefont {K.}~\bibnamefont {Semba}},\ }\bibfield  {title} {\enquote
  {\bibinfo {title} {Characteristic spectra of circuit quantum electrodynamics
  systems from the ultrastrong- to the deep-strong-coupling regime},}\ }\href
  {\doibase 10.1103/PhysRevA.95.053824} {\bibfield  {journal} {\bibinfo
  {journal} {Phys. Rev. A}\ }\textbf {\bibinfo {volume} {95}},\ \bibinfo
  {pages} {053824} (\bibinfo {year} {2017})}\BibitemShut {NoStop}%
\bibitem [{\citenamefont {Bosman}\ \emph {et~al.}(2017)\citenamefont {Bosman},
  \citenamefont {Gely}, \citenamefont {Singh}, \citenamefont {Bruno},
  \citenamefont {Bothner},\ and\ \citenamefont {Steele}}]{Bosman2017Njpqi}%
  \BibitemOpen
  \bibfield  {author} {\bibinfo {author} {\bibfnamefont {S.~J.}\ \bibnamefont
  {Bosman}}, \bibinfo {author} {\bibfnamefont {M.~F.}\ \bibnamefont {Gely}},
  \bibinfo {author} {\bibfnamefont {V.}~\bibnamefont {Singh}}, \bibinfo
  {author} {\bibfnamefont {A.}~\bibnamefont {Bruno}}, \bibinfo {author}
  {\bibfnamefont {D.}~\bibnamefont {Bothner}}, \ and\ \bibinfo {author}
  {\bibfnamefont {G.~A.}\ \bibnamefont {Steele}},\ }\bibfield  {title}
  {\enquote {\bibinfo {title} {Multi-mode ultra-strong coupling in circuit
  quantum electrodynamics},}\ }\href
  {https://doi.org/10.1038/s41534-017-0046-y} {\bibfield  {journal} {\bibinfo
  {journal} {npj Quant. Info.}\ }\textbf {\bibinfo {volume} {3}} (\bibinfo
  {year} {2017})}\BibitemShut {NoStop}%
\bibitem [{\citenamefont {Braum\"{u}ller}\ \emph {et~al.}(2017)\citenamefont
  {Braum\"{u}ller}, \citenamefont {Marthaler}, \citenamefont {Schneider},
  \citenamefont {Stehli}, \citenamefont {Rotzinger}, \citenamefont {Weides},\
  and\ \citenamefont {Ustinov}}]{Braumller2017NC}%
  \BibitemOpen
  \bibfield  {author} {\bibinfo {author} {\bibfnamefont {J.}~\bibnamefont
  {Braum\"{u}ller}}, \bibinfo {author} {\bibfnamefont {M.}~\bibnamefont
  {Marthaler}}, \bibinfo {author} {\bibfnamefont {A.}~\bibnamefont
  {Schneider}}, \bibinfo {author} {\bibfnamefont {A.}~\bibnamefont {Stehli}},
  \bibinfo {author} {\bibfnamefont {H.}~\bibnamefont {Rotzinger}}, \bibinfo
  {author} {\bibfnamefont {M.}~\bibnamefont {Weides}}, \ and\ \bibinfo {author}
  {\bibfnamefont {A.~V.}\ \bibnamefont {Ustinov}},\ }\bibfield  {title}
  {\enquote {\bibinfo {title} {Analog quantum simulation of the {R}abi model in
  the ultra-strong coupling regime},}\ }\href {\doibase
  10.1038/s41467-017-00894-w} {\bibfield  {journal} {\bibinfo  {journal} {Nat.
  Commun.}\ }\textbf {\bibinfo {volume} {8}},\ \bibinfo {pages} {779} (\bibinfo
  {year} {2017})}\BibitemShut {NoStop}%
\bibitem [{\citenamefont {Lv}\ \emph {et~al.}(2018)\citenamefont {Lv},
  \citenamefont {An}, \citenamefont {Liu}, \citenamefont {Zhang}, \citenamefont
  {Pedernales}, \citenamefont {Lamata}, \citenamefont {Solano},\ and\
  \citenamefont {Kim}}]{Lv2018PRX}%
  \BibitemOpen
  \bibfield  {author} {\bibinfo {author} {\bibfnamefont {D.}~\bibnamefont
  {Lv}}, \bibinfo {author} {\bibfnamefont {S.}~\bibnamefont {An}}, \bibinfo
  {author} {\bibfnamefont {Z.}~\bibnamefont {Liu}}, \bibinfo {author}
  {\bibfnamefont {J.-N.}\ \bibnamefont {Zhang}}, \bibinfo {author}
  {\bibfnamefont {J.~S.}\ \bibnamefont {Pedernales}}, \bibinfo {author}
  {\bibfnamefont {L.}~\bibnamefont {Lamata}}, \bibinfo {author} {\bibfnamefont
  {E.}~\bibnamefont {Solano}}, \ and\ \bibinfo {author} {\bibfnamefont
  {K.}~\bibnamefont {Kim}},\ }\bibfield  {title} {\enquote {\bibinfo {title}
  {Quantum simulation of the quantum {R}abi model in a trapped ion},}\ }\href
  {\doibase 10.1103/PhysRevX.8.021027} {\bibfield  {journal} {\bibinfo
  {journal} {Phys. Rev. X}\ }\textbf {\bibinfo {volume} {8}},\ \bibinfo {pages}
  {021027} (\bibinfo {year} {2018})}\BibitemShut {NoStop}%
\bibitem [{\citenamefont {Di~Stefano}\ \emph {et~al.}(2019)\citenamefont
  {Di~Stefano} \emph {et~al.}}]{DiStefano2019Np}%
  \BibitemOpen
  \bibfield  {author} {\bibinfo {author} {\bibfnamefont {O.}~\bibnamefont
  {Di~Stefano}} \emph {et~al.},\ }\bibfield  {title} {\enquote {\bibinfo
  {title} {Resolution of gauge ambiguities in ultrastrong-coupling cavity
  quantum electrodynamics},}\ }\href {\doibase 10.1038/s41567-019-0534-4}
  {\bibfield  {journal} {\bibinfo  {journal} {Nat. Phys.}\ }\textbf {\bibinfo
  {volume} {15}},\ \bibinfo {pages} {803–808} (\bibinfo {year}
  {2019})}\BibitemShut {NoStop}%
\bibitem [{\citenamefont {Cai}\ \emph {et~al.}(2021)\citenamefont {Cai},
  \citenamefont {Liu}, \citenamefont {Zhao}, \citenamefont {Wu}, \citenamefont
  {Mei}, \citenamefont {Jiang}, \citenamefont {He}, \citenamefont {Zhang},
  \citenamefont {Zhou},\ and\ \citenamefont {Duan}}]{Cai2021NC}%
  \BibitemOpen
  \bibfield  {author} {\bibinfo {author} {\bibfnamefont {M.-L.}\ \bibnamefont
  {Cai}}, \bibinfo {author} {\bibfnamefont {Z.-D.}\ \bibnamefont {Liu}},
  \bibinfo {author} {\bibfnamefont {W.-D.}\ \bibnamefont {Zhao}}, \bibinfo
  {author} {\bibfnamefont {Y.-K.}\ \bibnamefont {Wu}}, \bibinfo {author}
  {\bibfnamefont {Q.-X.}\ \bibnamefont {Mei}}, \bibinfo {author} {\bibfnamefont
  {Y.}~\bibnamefont {Jiang}}, \bibinfo {author} {\bibfnamefont
  {L.}~\bibnamefont {He}}, \bibinfo {author} {\bibfnamefont {X.}~\bibnamefont
  {Zhang}}, \bibinfo {author} {\bibfnamefont {Z.-C.}\ \bibnamefont {Zhou}}, \
  and\ \bibinfo {author} {\bibfnamefont {L.-M.}\ \bibnamefont {Duan}},\
  }\bibfield  {title} {\enquote {\bibinfo {title} {Observation of a quantum
  phase transition in the quantum {R}abi model with a single trapped ion},}\
  }\href {https://doi.org/10.1038/s41467-021-21425-8} {\bibfield  {journal}
  {\bibinfo  {journal} {Nat. Commun.}\ }\textbf {\bibinfo {volume} {12}},\
  \bibinfo {pages} {1126} (\bibinfo {year} {2021})}\BibitemShut {NoStop}%
\bibitem [{\citenamefont {Chen}\ \emph
  {et~al.}(2021{\natexlab{a}})\citenamefont {Chen}, \citenamefont {Wu},
  \citenamefont {Jiang}, \citenamefont {L\"{u}}, \citenamefont {Peng},\ and\
  \citenamefont {Du}}]{Chen2021NC}%
  \BibitemOpen
  \bibfield  {author} {\bibinfo {author} {\bibfnamefont {X.}~\bibnamefont
  {Chen}}, \bibinfo {author} {\bibfnamefont {Z.}~\bibnamefont {Wu}}, \bibinfo
  {author} {\bibfnamefont {M.}~\bibnamefont {Jiang}}, \bibinfo {author}
  {\bibfnamefont {X.-Y.}\ \bibnamefont {L\"{u}}}, \bibinfo {author}
  {\bibfnamefont {X.}~\bibnamefont {Peng}}, \ and\ \bibinfo {author}
  {\bibfnamefont {J.}~\bibnamefont {Du}},\ }\bibfield  {title} {\enquote
  {\bibinfo {title} {Experimental quantum simulation of superradiant phase
  transition beyond no-go theorem via antisqueezing},}\ }\href
  {https://doi.org/10.1038/s41467-021-26573-5} {\bibfield  {journal} {\bibinfo
  {journal} {Nat. Commun.}\ }\textbf {\bibinfo {volume} {12}},\ \bibinfo
  {pages} {6281} (\bibinfo {year} {2021}{\natexlab{a}})}\BibitemShut {NoStop}%
\bibitem [{\citenamefont {Zheng}\ \emph {et~al.}(2023)\citenamefont {Zheng}
  \emph {et~al.}}]{Zheng2023PRL}%
  \BibitemOpen
  \bibfield  {author} {\bibinfo {author} {\bibfnamefont {R.-H.}\ \bibnamefont
  {Zheng}} \emph {et~al.},\ }\bibfield  {title} {\enquote {\bibinfo {title}
  {Observation of a superradiant phase transition with emergent cat states},}\
  }\href {\doibase 10.1103/PhysRevLett.131.113601} {\bibfield  {journal}
  {\bibinfo  {journal} {Phys. Rev. Lett.}\ }\textbf {\bibinfo {volume} {131}},\
  \bibinfo {pages} {113601} (\bibinfo {year} {2023})}\BibitemShut {NoStop}%
\bibitem [{\citenamefont {Cao}\ \emph {et~al.}(2011)\citenamefont {Cao},
  \citenamefont {You}, \citenamefont {Zheng},\ and\ \citenamefont
  {Nori}}]{Cao2011Njp}%
  \BibitemOpen
  \bibfield  {author} {\bibinfo {author} {\bibfnamefont {X.}~\bibnamefont
  {Cao}}, \bibinfo {author} {\bibfnamefont {J.~Q.}\ \bibnamefont {You}},
  \bibinfo {author} {\bibfnamefont {H.}~\bibnamefont {Zheng}}, \ and\ \bibinfo
  {author} {\bibfnamefont {F.}~\bibnamefont {Nori}},\ }\bibfield  {title}
  {\enquote {\bibinfo {title} {A qubit strongly coupled to a resonant cavity:
  asymmetry of the spontaneous emission spectrum beyond the rotating wave
  approximation},}\ }\href {\doibase 10.1088/1367-2630/13/7/073002} {\bibfield
  {journal} {\bibinfo  {journal} {New J. Phys.}\ }\textbf {\bibinfo {volume}
  {13}},\ \bibinfo {pages} {073002} (\bibinfo {year} {2011})}\BibitemShut
  {NoStop}%
\bibitem [{\citenamefont {Ashhab}\ and\ \citenamefont
  {Nori}(2010)}]{Ashhab2010Pra}%
  \BibitemOpen
  \bibfield  {author} {\bibinfo {author} {\bibfnamefont {S.}~\bibnamefont
  {Ashhab}}\ and\ \bibinfo {author} {\bibfnamefont {F.}~\bibnamefont {Nori}},\
  }\bibfield  {title} {\enquote {\bibinfo {title} {Qubit-oscillator systems in
  the ultrastrong-coupling regime and their potential for preparing
  nonclassical states},}\ }\href {\doibase 10.1103/PhysRevA.81.042311}
  {\bibfield  {journal} {\bibinfo  {journal} {Phys. Rev. A}\ }\textbf {\bibinfo
  {volume} {81}},\ \bibinfo {pages} {042311} (\bibinfo {year}
  {2010})}\BibitemShut {NoStop}%
\bibitem [{\citenamefont {Ashhab}(2013)}]{Ashhab2013}%
  \BibitemOpen
  \bibfield  {author} {\bibinfo {author} {\bibfnamefont {S.}~\bibnamefont
  {Ashhab}},\ }\bibfield  {title} {\enquote {\bibinfo {title} {Superradiance
  transition in a system with a single qubit and a single oscillator},}\ }\href
  {\doibase 10.1103/PhysRevA.87.013826} {\bibfield  {journal} {\bibinfo
  {journal} {Phys. Rev. A}\ }\textbf {\bibinfo {volume} {87}},\ \bibinfo
  {pages} {013826} (\bibinfo {year} {2013})}\BibitemShut {NoStop}%
\bibitem [{\citenamefont {Chen}\ \emph
  {et~al.}(2021{\natexlab{b}})\citenamefont {Chen}, \citenamefont {Qin},
  \citenamefont {Wang}, \citenamefont {Miranowicz},\ and\ \citenamefont
  {Nori}}]{Chen2021}%
  \BibitemOpen
  \bibfield  {author} {\bibinfo {author} {\bibfnamefont {Y.-H.}\ \bibnamefont
  {Chen}}, \bibinfo {author} {\bibfnamefont {W.}~\bibnamefont {Qin}}, \bibinfo
  {author} {\bibfnamefont {X.}~\bibnamefont {Wang}}, \bibinfo {author}
  {\bibfnamefont {A.}~\bibnamefont {Miranowicz}}, \ and\ \bibinfo {author}
  {\bibfnamefont {F.}~\bibnamefont {Nori}},\ }\bibfield  {title} {\enquote
  {\bibinfo {title} {Shortcuts to adiabaticity for the quantum {R}abi model:
  Efficient generation of giant entangled cat states via parametric
  amplification},}\ }\href {\doibase 10.1103/PhysRevLett.126.023602} {\bibfield
   {journal} {\bibinfo  {journal} {Phys. Rev. Lett.}\ }\textbf {\bibinfo
  {volume} {126}},\ \bibinfo {pages} {023602} (\bibinfo {year}
  {2021}{\natexlab{b}})}\BibitemShut {NoStop}%
\bibitem [{\citenamefont {Chen}\ \emph
  {et~al.}(2024{\natexlab{a}})\citenamefont {Chen}, \citenamefont {Shi},
  \citenamefont {Nori},\ and\ \citenamefont {Xia}}]{Chen2024Prl}%
  \BibitemOpen
  \bibfield  {author} {\bibinfo {author} {\bibfnamefont {Y.-H.}\ \bibnamefont
  {Chen}}, \bibinfo {author} {\bibfnamefont {Z.-C.}\ \bibnamefont {Shi}},
  \bibinfo {author} {\bibfnamefont {F.}~\bibnamefont {Nori}}, \ and\ \bibinfo
  {author} {\bibfnamefont {Y.}~\bibnamefont {Xia}},\ }\bibfield  {title}
  {\enquote {\bibinfo {title} {Error-tolerant amplification and simulation of
  the ultrastrong-coupling quantum {R}abi model},}\ }\href {\doibase
  10.1103/PhysRevLett.133.033603} {\bibfield  {journal} {\bibinfo  {journal}
  {Phys. Rev. Lett.}\ }\textbf {\bibinfo {volume} {133}},\ \bibinfo {pages}
  {033603} (\bibinfo {year} {2024}{\natexlab{a}})}\BibitemShut {NoStop}%
\bibitem [{\citenamefont {Ridolfo}\ \emph {et~al.}(2012)\citenamefont
  {Ridolfo}, \citenamefont {Leib}, \citenamefont {Savasta},\ and\ \citenamefont
  {Hartmann}}]{Ridolfo2012}%
  \BibitemOpen
  \bibfield  {author} {\bibinfo {author} {\bibfnamefont {A.}~\bibnamefont
  {Ridolfo}}, \bibinfo {author} {\bibfnamefont {M.}~\bibnamefont {Leib}},
  \bibinfo {author} {\bibfnamefont {S.}~\bibnamefont {Savasta}}, \ and\
  \bibinfo {author} {\bibfnamefont {M.~J.}\ \bibnamefont {Hartmann}},\
  }\bibfield  {title} {\enquote {\bibinfo {title} {Photon blockade in the
  ultrastrong coupling regime},}\ }\href {\doibase
  10.1103/PhysRevLett.109.193602} {\bibfield  {journal} {\bibinfo  {journal}
  {Phys. Rev. Lett.}\ }\textbf {\bibinfo {volume} {109}},\ \bibinfo {pages}
  {193602} (\bibinfo {year} {2012})}\BibitemShut {NoStop}%
\bibitem [{\citenamefont {Ridolfo}\ \emph {et~al.}(2013)\citenamefont
  {Ridolfo}, \citenamefont {Savasta},\ and\ \citenamefont
  {Hartmann}}]{Ridolfo2013}%
  \BibitemOpen
  \bibfield  {author} {\bibinfo {author} {\bibfnamefont {A.}~\bibnamefont
  {Ridolfo}}, \bibinfo {author} {\bibfnamefont {S.}~\bibnamefont {Savasta}}, \
  and\ \bibinfo {author} {\bibfnamefont {M.~J.}\ \bibnamefont {Hartmann}},\
  }\bibfield  {title} {\enquote {\bibinfo {title} {Nonclassical radiation from
  thermal cavities in the ultrastrong coupling regime},}\ }\href {\doibase
  10.1103/PhysRevLett.110.163601} {\bibfield  {journal} {\bibinfo  {journal}
  {Phys. Rev. Lett.}\ }\textbf {\bibinfo {volume} {110}},\ \bibinfo {pages}
  {163601} (\bibinfo {year} {2013})}\BibitemShut {NoStop}%
\bibitem [{\citenamefont {Hwang}\ \emph {et~al.}(2015)\citenamefont {Hwang},
  \citenamefont {Puebla},\ and\ \citenamefont {Plenio}}]{Hwang2015Prl}%
  \BibitemOpen
  \bibfield  {author} {\bibinfo {author} {\bibfnamefont {M.-J.}\ \bibnamefont
  {Hwang}}, \bibinfo {author} {\bibfnamefont {R.}~\bibnamefont {Puebla}}, \
  and\ \bibinfo {author} {\bibfnamefont {M.~B.}\ \bibnamefont {Plenio}},\
  }\bibfield  {title} {\enquote {\bibinfo {title} {Quantum phase transition and
  universal dynamics in the {R}abi model},}\ }\href {\doibase
  10.1103/PhysRevLett.115.180404} {\bibfield  {journal} {\bibinfo  {journal}
  {Phys. Rev. Lett.}\ }\textbf {\bibinfo {volume} {115}},\ \bibinfo {pages}
  {180404} (\bibinfo {year} {2015})}\BibitemShut {NoStop}%
\bibitem [{\citenamefont {Shammah}\ \emph {et~al.}(2017)\citenamefont
  {Shammah}, \citenamefont {Lambert}, \citenamefont {Nori},\ and\ \citenamefont
  {De~Liberato}}]{Shammah2017}%
  \BibitemOpen
  \bibfield  {author} {\bibinfo {author} {\bibfnamefont {N.}~\bibnamefont
  {Shammah}}, \bibinfo {author} {\bibfnamefont {N.}~\bibnamefont {Lambert}},
  \bibinfo {author} {\bibfnamefont {F.}~\bibnamefont {Nori}}, \ and\ \bibinfo
  {author} {\bibfnamefont {S.}~\bibnamefont {De~Liberato}},\ }\bibfield
  {title} {\enquote {\bibinfo {title} {Superradiance with local phase-breaking
  effects},}\ }\href {\doibase 10.1103/PhysRevA.96.023863} {\bibfield
  {journal} {\bibinfo  {journal} {Phys. Rev. A}\ }\textbf {\bibinfo {volume}
  {96}},\ \bibinfo {pages} {023863} (\bibinfo {year} {2017})}\BibitemShut
  {NoStop}%
\bibitem [{\citenamefont {Shapira}\ \emph {et~al.}(2018)\citenamefont
  {Shapira}, \citenamefont {Shaniv}, \citenamefont {Manovitz}, \citenamefont
  {Akerman},\ and\ \citenamefont {Ozeri}}]{Shapira2018}%
  \BibitemOpen
  \bibfield  {author} {\bibinfo {author} {\bibfnamefont {Y.}~\bibnamefont
  {Shapira}}, \bibinfo {author} {\bibfnamefont {R.}~\bibnamefont {Shaniv}},
  \bibinfo {author} {\bibfnamefont {T.}~\bibnamefont {Manovitz}}, \bibinfo
  {author} {\bibfnamefont {N.}~\bibnamefont {Akerman}}, \ and\ \bibinfo
  {author} {\bibfnamefont {R.}~\bibnamefont {Ozeri}},\ }\bibfield  {title}
  {\enquote {\bibinfo {title} {Robust entanglement gates for trapped-ion
  qubits},}\ }\href {\doibase 10.1103/PhysRevLett.121.180502} {\bibfield
  {journal} {\bibinfo  {journal} {Phys. Rev. Lett.}\ }\textbf {\bibinfo
  {volume} {121}},\ \bibinfo {pages} {180502} (\bibinfo {year}
  {2018})}\BibitemShut {NoStop}%
\bibitem [{\citenamefont {Chen}\ \emph
  {et~al.}(2024{\natexlab{b}})\citenamefont {Chen} \emph
  {et~al.}}]{Chen2024CP}%
  \BibitemOpen
  \bibfield  {author} {\bibinfo {author} {\bibfnamefont {Y.-H.}\ \bibnamefont
  {Chen}} \emph {et~al.},\ }\bibfield  {title} {\enquote {\bibinfo {title}
  {Sudden change of the photon output field marks phase transitions in the
  quantum {R}abi model},}\ }\href
  {http://dx.doi.org/10.1038/s42005-023-01457-w} {\bibfield  {journal}
  {\bibinfo  {journal} {Commun. Phys.}\ }\textbf {\bibinfo {volume} {7}},\
  \bibinfo {pages} {5} (\bibinfo {year} {2024}{\natexlab{b}})}\BibitemShut
  {NoStop}%
\bibitem [{\citenamefont {Wang}\ \emph {et~al.}(2014)\citenamefont {Wang} \emph
  {et~al.}}]{Wang2014Njp}%
  \BibitemOpen
  \bibfield  {author} {\bibinfo {author} {\bibfnamefont {T.-L.}\ \bibnamefont
  {Wang}} \emph {et~al.},\ }\bibfield  {title} {\enquote {\bibinfo {title}
  {Quantum {F}isher information as a signature of the superradiant quantum
  phase transition},}\ }\href {\doibase 10.1088/1367-2630/16/6/063039}
  {\bibfield  {journal} {\bibinfo  {journal} {New Journal of Physics}\ }\textbf
  {\bibinfo {volume} {16}},\ \bibinfo {pages} {063039} (\bibinfo {year}
  {2014})}\BibitemShut {NoStop}%
\bibitem [{\citenamefont {Chu}\ \emph {et~al.}(2021)\citenamefont {Chu},
  \citenamefont {Zhang}, \citenamefont {Yu},\ and\ \citenamefont
  {Cai}}]{Chu2021Prl}%
  \BibitemOpen
  \bibfield  {author} {\bibinfo {author} {\bibfnamefont {Y.}~\bibnamefont
  {Chu}}, \bibinfo {author} {\bibfnamefont {S.}~\bibnamefont {Zhang}}, \bibinfo
  {author} {\bibfnamefont {B.}~\bibnamefont {Yu}}, \ and\ \bibinfo {author}
  {\bibfnamefont {J.}~\bibnamefont {Cai}},\ }\bibfield  {title} {\enquote
  {\bibinfo {title} {Dynamic framework for criticality-enhanced quantum
  sensing},}\ }\href {\doibase 10.1103/PhysRevLett.126.010502} {\bibfield
  {journal} {\bibinfo  {journal} {Phys. Rev. Lett.}\ }\textbf {\bibinfo
  {volume} {126}},\ \bibinfo {pages} {010502} (\bibinfo {year}
  {2021})}\BibitemShut {NoStop}%
\bibitem [{\citenamefont {Garbe}\ \emph {et~al.}(2020)\citenamefont {Garbe},
  \citenamefont {Bina}, \citenamefont {Keller}, \citenamefont {Paris},\ and\
  \citenamefont {Felicetti}}]{Garbe2020Prl}%
  \BibitemOpen
  \bibfield  {author} {\bibinfo {author} {\bibfnamefont {L.}~\bibnamefont
  {Garbe}}, \bibinfo {author} {\bibfnamefont {M.}~\bibnamefont {Bina}},
  \bibinfo {author} {\bibfnamefont {A.}~\bibnamefont {Keller}}, \bibinfo
  {author} {\bibfnamefont {M.~G.~A.}\ \bibnamefont {Paris}}, \ and\ \bibinfo
  {author} {\bibfnamefont {S.}~\bibnamefont {Felicetti}},\ }\bibfield  {title}
  {\enquote {\bibinfo {title} {Critical quantum metrology with a
  finite-component quantum phase transition},}\ }\href {\doibase
  10.1103/PhysRevLett.124.120504} {\bibfield  {journal} {\bibinfo  {journal}
  {Phys. Rev. Lett.}\ }\textbf {\bibinfo {volume} {124}},\ \bibinfo {pages}
  {120504} (\bibinfo {year} {2020})}\BibitemShut {NoStop}%
\bibitem [{\citenamefont {Xu}\ \emph {et~al.}(2020)\citenamefont {Xu} \emph
  {et~al.}}]{Xu2020Sa}%
  \BibitemOpen
  \bibfield  {author} {\bibinfo {author} {\bibfnamefont {K.}~\bibnamefont {Xu}}
  \emph {et~al.},\ }\bibfield  {title} {\enquote {\bibinfo {title} {Probing
  dynamical phase transitions with a superconducting quantum simulator},}\
  }\href {\doibase 10.1126/sciadv.aba4935} {\bibfield  {journal} {\bibinfo
  {journal} {Science Advances}\ }\textbf {\bibinfo {volume} {6}},\ \bibinfo
  {pages} {eaba4935} (\bibinfo {year} {2020})}\BibitemShut {NoStop}%
\bibitem [{\citenamefont {Hotter}\ \emph {et~al.}(2024)\citenamefont {Hotter},
  \citenamefont {Ritsch},\ and\ \citenamefont {Gietka}}]{Hotter2024Prl}%
  \BibitemOpen
  \bibfield  {author} {\bibinfo {author} {\bibfnamefont {C.}~\bibnamefont
  {Hotter}}, \bibinfo {author} {\bibfnamefont {H.}~\bibnamefont {Ritsch}}, \
  and\ \bibinfo {author} {\bibfnamefont {K.}~\bibnamefont {Gietka}},\
  }\bibfield  {title} {\enquote {\bibinfo {title} {Combining critical and
  quantum metrology},}\ }\href {\doibase 10.1103/PhysRevLett.132.060801}
  {\bibfield  {journal} {\bibinfo  {journal} {Phys. Rev. Lett.}\ }\textbf
  {\bibinfo {volume} {132}},\ \bibinfo {pages} {060801} (\bibinfo {year}
  {2024})}\BibitemShut {NoStop}%
\bibitem [{\citenamefont {Wan}\ \emph {et~al.}(2024)\citenamefont {Wan},
  \citenamefont {Shi},\ and\ \citenamefont {Guan}}]{Wan2024Prb}%
  \BibitemOpen
  \bibfield  {author} {\bibinfo {author} {\bibfnamefont {Q.-K.}\ \bibnamefont
  {Wan}}, \bibinfo {author} {\bibfnamefont {H.-L.}\ \bibnamefont {Shi}}, \ and\
  \bibinfo {author} {\bibfnamefont {X.-W.}\ \bibnamefont {Guan}},\ }\bibfield
  {title} {\enquote {\bibinfo {title} {Quantum-enhanced metrology in cavity
  magnonics},}\ }\href {\doibase 10.1103/PhysRevB.109.L041301} {\bibfield
  {journal} {\bibinfo  {journal} {Phys. Rev. B}\ }\textbf {\bibinfo {volume}
  {109}},\ \bibinfo {pages} {L041301} (\bibinfo {year} {2024})}\BibitemShut
  {NoStop}%
\bibitem [{\citenamefont {Zhu}\ \emph {et~al.}(2024)\citenamefont {Zhu},
  \citenamefont {L\"u}, \citenamefont {Ning}, \citenamefont {Shen},
  \citenamefont {Wu},\ and\ \citenamefont {Yang}}]{Zhu2024Pra}%
  \BibitemOpen
  \bibfield  {author} {\bibinfo {author} {\bibfnamefont {X.}~\bibnamefont
  {Zhu}}, \bibinfo {author} {\bibfnamefont {J.-H.}\ \bibnamefont {L\"u}},
  \bibinfo {author} {\bibfnamefont {W.}~\bibnamefont {Ning}}, \bibinfo {author}
  {\bibfnamefont {L.-T.}\ \bibnamefont {Shen}}, \bibinfo {author}
  {\bibfnamefont {F.}~\bibnamefont {Wu}}, \ and\ \bibinfo {author}
  {\bibfnamefont {Z.-B.}\ \bibnamefont {Yang}},\ }\bibfield  {title} {\enquote
  {\bibinfo {title} {Quantum geometric tensor and critical metrology in the
  anisotropic {D}icke model},}\ }\href {\doibase 10.1103/PhysRevA.109.052621}
  {\bibfield  {journal} {\bibinfo  {journal} {Phys. Rev. A}\ }\textbf {\bibinfo
  {volume} {109}},\ \bibinfo {pages} {052621} (\bibinfo {year}
  {2024})}\BibitemShut {NoStop}%
\bibitem [{\citenamefont {Shammah}\ \emph {et~al.}(2018)\citenamefont {Shammah}
  \emph {et~al.}}]{Shammah2018Pra}%
  \BibitemOpen
  \bibfield  {author} {\bibinfo {author} {\bibfnamefont {N.}~\bibnamefont
  {Shammah}} \emph {et~al.},\ }\bibfield  {title} {\enquote {\bibinfo {title}
  {Open quantum systems with local and collective incoherent processes:
  Efficient numerical simulations using permutational invariance},}\ }\href
  {\doibase 10.1103/PhysRevA.98.063815} {\bibfield  {journal} {\bibinfo
  {journal} {Phys. Rev. A}\ }\textbf {\bibinfo {volume} {98}},\ \bibinfo
  {pages} {063815} (\bibinfo {year} {2018})}\BibitemShut {NoStop}%
\bibitem [{\citenamefont {De~Liberato}\ \emph {et~al.}(2009)\citenamefont
  {De~Liberato}, \citenamefont {Gerace}, \citenamefont {Carusotto},\ and\
  \citenamefont {Ciuti}}]{DeLiberato2009Pra}%
  \BibitemOpen
  \bibfield  {author} {\bibinfo {author} {\bibfnamefont {S.}~\bibnamefont
  {De~Liberato}}, \bibinfo {author} {\bibfnamefont {D.}~\bibnamefont {Gerace}},
  \bibinfo {author} {\bibfnamefont {I.}~\bibnamefont {Carusotto}}, \ and\
  \bibinfo {author} {\bibfnamefont {C.}~\bibnamefont {Ciuti}},\ }\bibfield
  {title} {\enquote {\bibinfo {title} {Extracavity quantum vacuum radiation
  from a single qubit},}\ }\href {\doibase 10.1103/PhysRevA.80.053810}
  {\bibfield  {journal} {\bibinfo  {journal} {Phys. Rev. A}\ }\textbf {\bibinfo
  {volume} {80}},\ \bibinfo {pages} {053810} (\bibinfo {year}
  {2009})}\BibitemShut {NoStop}%
\bibitem [{\citenamefont {Beaudoin}\ \emph {et~al.}(2011)\citenamefont
  {Beaudoin}, \citenamefont {Gambetta},\ and\ \citenamefont
  {Blais}}]{Beaudoin2011Pra}%
  \BibitemOpen
  \bibfield  {author} {\bibinfo {author} {\bibfnamefont {F.}~\bibnamefont
  {Beaudoin}}, \bibinfo {author} {\bibfnamefont {J.~M.}\ \bibnamefont
  {Gambetta}}, \ and\ \bibinfo {author} {\bibfnamefont {A.}~\bibnamefont
  {Blais}},\ }\bibfield  {title} {\enquote {\bibinfo {title} {Dissipation and
  ultrastrong coupling in circuit {QED}},}\ }\href {\doibase
  10.1103/PhysRevA.84.043832} {\bibfield  {journal} {\bibinfo  {journal} {Phys.
  Rev. A}\ }\textbf {\bibinfo {volume} {84}},\ \bibinfo {pages} {043832}
  (\bibinfo {year} {2011})}\BibitemShut {NoStop}%
\bibitem [{\citenamefont {Settineri}\ \emph {et~al.}(2018)\citenamefont
  {Settineri} \emph {et~al.}}]{Settineri2018Pra}%
  \BibitemOpen
  \bibfield  {author} {\bibinfo {author} {\bibfnamefont {A.}~\bibnamefont
  {Settineri}} \emph {et~al.},\ }\bibfield  {title} {\enquote {\bibinfo {title}
  {Dissipation and thermal noise in hybrid quantum systems in the
  ultrastrong-coupling regime},}\ }\href {\doibase 10.1103/PhysRevA.98.053834}
  {\bibfield  {journal} {\bibinfo  {journal} {Phys. Rev. A}\ }\textbf {\bibinfo
  {volume} {98}},\ \bibinfo {pages} {053834} (\bibinfo {year}
  {2018})}\BibitemShut {NoStop}%
\bibitem [{\citenamefont {Mercurio}\ \emph {et~al.}(2023)\citenamefont
  {Mercurio}, \citenamefont {Abo}, \citenamefont {Mauceri}, \citenamefont
  {Russo}, \citenamefont {Macr\`{\i}}, \citenamefont {Miranowicz},
  \citenamefont {Savasta},\ and\ \citenamefont {Di~Stefano}}]{Mercurio2023Prl}%
  \BibitemOpen
  \bibfield  {author} {\bibinfo {author} {\bibfnamefont {A.}~\bibnamefont
  {Mercurio}}, \bibinfo {author} {\bibfnamefont {S.}~\bibnamefont {Abo}},
  \bibinfo {author} {\bibfnamefont {F.}~\bibnamefont {Mauceri}}, \bibinfo
  {author} {\bibfnamefont {E.}~\bibnamefont {Russo}}, \bibinfo {author}
  {\bibfnamefont {V.}~\bibnamefont {Macr\`{\i}}}, \bibinfo {author}
  {\bibfnamefont {A.}~\bibnamefont {Miranowicz}}, \bibinfo {author}
  {\bibfnamefont {S.}~\bibnamefont {Savasta}}, \ and\ \bibinfo {author}
  {\bibfnamefont {O.}~\bibnamefont {Di~Stefano}},\ }\bibfield  {title}
  {\enquote {\bibinfo {title} {Pure dephasing of light-matter systems in the
  ultrastrong and deep-strong coupling regimes},}\ }\href {\doibase
  10.1103/PhysRevLett.130.123601} {\bibfield  {journal} {\bibinfo  {journal}
  {Phys. Rev. Lett.}\ }\textbf {\bibinfo {volume} {130}},\ \bibinfo {pages}
  {123601} (\bibinfo {year} {2023})}\BibitemShut {NoStop}%
\bibitem [{\citenamefont {Breuer}\ and\ \citenamefont
  {Petruccione}(2002)}]{Breuer2002Book}%
  \BibitemOpen
  \bibfield  {author} {\bibinfo {author} {\bibfnamefont {H.~P.}\ \bibnamefont
  {Breuer}}\ and\ \bibinfo {author} {\bibfnamefont {F.}~\bibnamefont
  {Petruccione}},\ }\href
  {https://doi.org/10.1093/acprof:oso/9780199213900.001.0001} {\emph {\bibinfo
  {title} {The Theory of Open Quantum Systems}}}\ (\bibinfo  {publisher}
  {Oxford University Press},\ \bibinfo {address} {Oxford, England},\ \bibinfo
  {year} {2002})\BibitemShut {NoStop}%
\bibitem [{\citenamefont {Garziano}\ \emph {et~al.}(2013)\citenamefont
  {Garziano}, \citenamefont {Ridolfo}, \citenamefont {Stassi}, \citenamefont
  {Di~Stefano},\ and\ \citenamefont {Savasta}}]{Garziano2013Pra}%
  \BibitemOpen
  \bibfield  {author} {\bibinfo {author} {\bibfnamefont {L.}~\bibnamefont
  {Garziano}}, \bibinfo {author} {\bibfnamefont {A.}~\bibnamefont {Ridolfo}},
  \bibinfo {author} {\bibfnamefont {R.}~\bibnamefont {Stassi}}, \bibinfo
  {author} {\bibfnamefont {O.}~\bibnamefont {Di~Stefano}}, \ and\ \bibinfo
  {author} {\bibfnamefont {S.}~\bibnamefont {Savasta}},\ }\bibfield  {title}
  {\enquote {\bibinfo {title} {Switching on and off of ultrastrong light-matter
  interaction: Photon statistics of quantum vacuum radiation},}\ }\href
  {\doibase 10.1103/PhysRevA.88.063829} {\bibfield  {journal} {\bibinfo
  {journal} {Phys. Rev. A}\ }\textbf {\bibinfo {volume} {88}},\ \bibinfo
  {pages} {063829} (\bibinfo {year} {2013})}\BibitemShut {NoStop}%
\bibitem [{\citenamefont {Stassi}\ \emph {et~al.}(2013)\citenamefont {Stassi},
  \citenamefont {Ridolfo}, \citenamefont {Di~Stefano}, \citenamefont
  {Hartmann},\ and\ \citenamefont {Savasta}}]{Stassi2013Prl}%
  \BibitemOpen
  \bibfield  {author} {\bibinfo {author} {\bibfnamefont {R.}~\bibnamefont
  {Stassi}}, \bibinfo {author} {\bibfnamefont {A.}~\bibnamefont {Ridolfo}},
  \bibinfo {author} {\bibfnamefont {O.}~\bibnamefont {Di~Stefano}}, \bibinfo
  {author} {\bibfnamefont {M.~J.}\ \bibnamefont {Hartmann}}, \ and\ \bibinfo
  {author} {\bibfnamefont {S.}~\bibnamefont {Savasta}},\ }\bibfield  {title}
  {\enquote {\bibinfo {title} {Spontaneous conversion from virtual to real
  photons in the ultrastrong-coupling regime},}\ }\href {\doibase
  10.1103/PhysRevLett.110.243601} {\bibfield  {journal} {\bibinfo  {journal}
  {Phys. Rev. Lett.}\ }\textbf {\bibinfo {volume} {110}},\ \bibinfo {pages}
  {243601} (\bibinfo {year} {2013})}\BibitemShut {NoStop}%
\bibitem [{\citenamefont {Rossatto}\ \emph {et~al.}(2017)\citenamefont
  {Rossatto}, \citenamefont {Villas-B\^oas}, \citenamefont {Sanz},\ and\
  \citenamefont {Solano}}]{Rossatto2017Pra}%
  \BibitemOpen
  \bibfield  {author} {\bibinfo {author} {\bibfnamefont {D.~Z.}\ \bibnamefont
  {Rossatto}}, \bibinfo {author} {\bibfnamefont {C.~J.}\ \bibnamefont
  {Villas-B\^oas}}, \bibinfo {author} {\bibfnamefont {M.}~\bibnamefont {Sanz}},
  \ and\ \bibinfo {author} {\bibfnamefont {E.}~\bibnamefont {Solano}},\
  }\bibfield  {title} {\enquote {\bibinfo {title} {Spectral classification of
  coupling regimes in the quantum {R}abi model},}\ }\href {\doibase
  10.1103/PhysRevA.96.013849} {\bibfield  {journal} {\bibinfo  {journal} {Phys.
  Rev. A}\ }\textbf {\bibinfo {volume} {96}},\ \bibinfo {pages} {013849}
  (\bibinfo {year} {2017})}\BibitemShut {NoStop}%
\bibitem [{\citenamefont {De~Bernardis}(2023)}]{DeBernardis2023Pra}%
  \BibitemOpen
  \bibfield  {author} {\bibinfo {author} {\bibfnamefont {Daniele}\ \bibnamefont
  {De~Bernardis}},\ }\bibfield  {title} {\enquote {\bibinfo {title} {Relaxation
  breakdown and resonant tunneling in ultrastrong-coupling cavity {QED}},}\
  }\href {\doibase 10.1103/PhysRevA.108.043717} {\bibfield  {journal} {\bibinfo
   {journal} {Phys. Rev. A}\ }\textbf {\bibinfo {volume} {108}},\ \bibinfo
  {pages} {043717} (\bibinfo {year} {2023})}\BibitemShut {NoStop}%
\bibitem [{\citenamefont {Hu}\ \emph {et~al.}(2015)\citenamefont {Hu},
  \citenamefont {Huang}, \citenamefont {Liao}, \citenamefont {Tian},\ and\
  \citenamefont {Goan}}]{Hu2015Pra}%
  \BibitemOpen
  \bibfield  {author} {\bibinfo {author} {\bibfnamefont {D.}~\bibnamefont
  {Hu}}, \bibinfo {author} {\bibfnamefont {S.-Y.}\ \bibnamefont {Huang}},
  \bibinfo {author} {\bibfnamefont {J.-Q.}\ \bibnamefont {Liao}}, \bibinfo
  {author} {\bibfnamefont {L.}~\bibnamefont {Tian}}, \ and\ \bibinfo {author}
  {\bibfnamefont {H.-S.}\ \bibnamefont {Goan}},\ }\bibfield  {title} {\enquote
  {\bibinfo {title} {Quantum coherence in ultrastrong optomechanics},}\ }\href
  {\doibase 10.1103/PhysRevA.91.013812} {\bibfield  {journal} {\bibinfo
  {journal} {Phys. Rev. A}\ }\textbf {\bibinfo {volume} {91}},\ \bibinfo
  {pages} {013812} (\bibinfo {year} {2015})}\BibitemShut {NoStop}%
\bibitem [{\citenamefont {Giovannetti}\ \emph {et~al.}(2011)\citenamefont
  {Giovannetti}, \citenamefont {Lloyd},\ and\ \citenamefont
  {Maccone}}]{Giovannetti2011Nat}%
  \BibitemOpen
  \bibfield  {author} {\bibinfo {author} {\bibfnamefont {V.}~\bibnamefont
  {Giovannetti}}, \bibinfo {author} {\bibfnamefont {S.}~\bibnamefont {Lloyd}},
  \ and\ \bibinfo {author} {\bibfnamefont {L.}~\bibnamefont {Maccone}},\
  }\bibfield  {title} {\enquote {\bibinfo {title} {Advances in quantum
  metrology},}\ }\href {\doibase 10.1038/nphoton.2011.35} {\bibfield  {journal}
  {\bibinfo  {journal} {Nat. Photon.}\ }\textbf {\bibinfo {volume} {5}},\
  \bibinfo {pages} {222–229} (\bibinfo {year} {2011})}\BibitemShut {NoStop}%
\bibitem [{\citenamefont {Genoni}\ \emph {et~al.}(2011)\citenamefont {Genoni},
  \citenamefont {Olivares},\ and\ \citenamefont {Paris}}]{Genoni2011Prl}%
  \BibitemOpen
  \bibfield  {author} {\bibinfo {author} {\bibfnamefont {M.~G.}\ \bibnamefont
  {Genoni}}, \bibinfo {author} {\bibfnamefont {S.}~\bibnamefont {Olivares}}, \
  and\ \bibinfo {author} {\bibfnamefont {M.~G.~A.}\ \bibnamefont {Paris}},\
  }\bibfield  {title} {\enquote {\bibinfo {title} {Optical phase estimation in
  the presence of phase diffusion},}\ }\href {\doibase
  10.1103/PhysRevLett.106.153603} {\bibfield  {journal} {\bibinfo  {journal}
  {Phys. Rev. Lett.}\ }\textbf {\bibinfo {volume} {106}},\ \bibinfo {pages}
  {153603} (\bibinfo {year} {2011})}\BibitemShut {NoStop}%
\bibitem [{\citenamefont {Zhang}\ and\ \citenamefont
  {Fan}(2014)}]{Zhang2014Pra}%
  \BibitemOpen
  \bibfield  {author} {\bibinfo {author} {\bibfnamefont {Y.-R.}\ \bibnamefont
  {Zhang}}\ and\ \bibinfo {author} {\bibfnamefont {H.}~\bibnamefont {Fan}},\
  }\bibfield  {title} {\enquote {\bibinfo {title} {Quantum metrological bounds
  for vector parameters},}\ }\href {\doibase 10.1103/PhysRevA.90.043818}
  {\bibfield  {journal} {\bibinfo  {journal} {Phys. Rev. A}\ }\textbf {\bibinfo
  {volume} {90}},\ \bibinfo {pages} {043818} (\bibinfo {year}
  {2014})}\BibitemShut {NoStop}%
\bibitem [{\citenamefont {Liu}\ \emph {et~al.}(2015)\citenamefont {Liu},
  \citenamefont {Zhang}, \citenamefont {Chang}, \citenamefont {Yue},
  \citenamefont {Fan},\ and\ \citenamefont {Pan}}]{Liu2015Nc}%
  \BibitemOpen
  \bibfield  {author} {\bibinfo {author} {\bibfnamefont {G.-Q.}\ \bibnamefont
  {Liu}}, \bibinfo {author} {\bibfnamefont {Y.-R.}\ \bibnamefont {Zhang}},
  \bibinfo {author} {\bibfnamefont {Y.-C.}\ \bibnamefont {Chang}}, \bibinfo
  {author} {\bibfnamefont {J.-D.}\ \bibnamefont {Yue}}, \bibinfo {author}
  {\bibfnamefont {H.}~\bibnamefont {Fan}}, \ and\ \bibinfo {author}
  {\bibfnamefont {X.-Y.}\ \bibnamefont {Pan}},\ }\bibfield  {title} {\enquote
  {\bibinfo {title} {Demonstration of entanglement-enhanced phase estimation in
  solid},}\ }\href {http://dx.doi.org/10.1038/ncomms7726} {\bibfield  {journal}
  {\bibinfo  {journal} {Nature Communications}\ }\textbf {\bibinfo {volume}
  {6}},\ \bibinfo {pages} {6726} (\bibinfo {year} {2015})}\BibitemShut
  {NoStop}%
\bibitem [{\citenamefont {Cram\'{e}r}(1946)}]{Cramer1946book}%
  \BibitemOpen
  \bibfield  {author} {\bibinfo {author} {\bibfnamefont {H.}~\bibnamefont
  {Cram\'{e}r}},\ }\href {https://web.stanford.edu/~boyd/cvxbook/} {\emph
  {\bibinfo {title} {Mathematical Methods of Statistics}}}\ (\bibinfo
  {publisher} {Princeton University},\ \bibinfo {address} {Princeton, USA},\
  \bibinfo {year} {1946})\BibitemShut {NoStop}%
\bibitem [{\citenamefont {Rao}(1973)}]{Rao1973book}%
  \BibitemOpen
  \bibfield  {author} {\bibinfo {author} {\bibfnamefont {C.~R.}\ \bibnamefont
  {Rao}},\ }\href {\doibase 10.1002/9780470316436} {\emph {\bibinfo {title}
  {Linear Statistical Inference and its Applications}}}\ (\bibinfo  {publisher}
  {John Wiley \& Sons},\ \bibinfo {address} {New York, USA},\ \bibinfo {year}
  {1973})\BibitemShut {NoStop}%
\bibitem [{\citenamefont {Holevo}(2011)}]{Holevo2011Book}%
  \BibitemOpen
  \bibfield  {author} {\bibinfo {author} {\bibfnamefont {A.}~\bibnamefont
  {Holevo}},\ }\href {http://dx.doi.org/10.1007/978-88-7642-378-9} {\emph
  {\bibinfo {title} {Probabilistic and Statistical Aspects of Quantum
  Theory}}}\ (\bibinfo  {publisher} {Edizioni della Normale},\ \bibinfo
  {address} {Pisa, Italy},\ \bibinfo {year} {2011})\BibitemShut {NoStop}%
\bibitem [{\citenamefont {Ma}\ \emph {et~al.}(2011)\citenamefont {Ma},
  \citenamefont {Wang}, \citenamefont {Sun},\ and\ \citenamefont
  {Nori}}]{Ma2011PR}%
  \BibitemOpen
  \bibfield  {author} {\bibinfo {author} {\bibfnamefont {J.}~\bibnamefont
  {Ma}}, \bibinfo {author} {\bibfnamefont {X.}~\bibnamefont {Wang}}, \bibinfo
  {author} {\bibfnamefont {C.-P.}\ \bibnamefont {Sun}}, \ and\ \bibinfo
  {author} {\bibfnamefont {F.}~\bibnamefont {Nori}},\ }\bibfield  {title}
  {\enquote {\bibinfo {title} {Quantum spin squeezing},}\ }\href {\doibase
  10.1016/j.physrep.2011.08.003} {\bibfield  {journal} {\bibinfo  {journal}
  {Physics Reports}\ }\textbf {\bibinfo {volume} {509}},\ \bibinfo {pages}
  {89–165} (\bibinfo {year} {2011})}\BibitemShut {NoStop}%
\bibitem [{\citenamefont {Degen}\ \emph {et~al.}(2017)\citenamefont {Degen},
  \citenamefont {Reinhard},\ and\ \citenamefont {Cappellaro}}]{Degen2017Rmp}%
  \BibitemOpen
  \bibfield  {author} {\bibinfo {author} {\bibfnamefont {C.~L.}\ \bibnamefont
  {Degen}}, \bibinfo {author} {\bibfnamefont {F.}~\bibnamefont {Reinhard}}, \
  and\ \bibinfo {author} {\bibfnamefont {P.}~\bibnamefont {Cappellaro}},\
  }\bibfield  {title} {\enquote {\bibinfo {title} {Quantum sensing},}\ }\href
  {\doibase 10.1103/RevModPhys.89.035002} {\bibfield  {journal} {\bibinfo
  {journal} {Rev. Mod. Phys.}\ }\textbf {\bibinfo {volume} {89}},\ \bibinfo
  {pages} {035002} (\bibinfo {year} {2017})}\BibitemShut {NoStop}%
\bibitem [{\citenamefont {Lu}\ and\ \citenamefont {Wang}(2021)}]{Lu2021Prl}%
  \BibitemOpen
  \bibfield  {author} {\bibinfo {author} {\bibfnamefont {X.-M.}\ \bibnamefont
  {Lu}}\ and\ \bibinfo {author} {\bibfnamefont {X.}~\bibnamefont {Wang}},\
  }\bibfield  {title} {\enquote {\bibinfo {title} {Incorporating heisenberg's
  uncertainty principle into quantum multiparameter estimation},}\ }\href
  {\doibase 10.1103/PhysRevLett.126.120503} {\bibfield  {journal} {\bibinfo
  {journal} {Phys. Rev. Lett.}\ }\textbf {\bibinfo {volume} {126}},\ \bibinfo
  {pages} {120503} (\bibinfo {year} {2021})}\BibitemShut {NoStop}%
\bibitem [{\citenamefont {Xu}\ \emph {et~al.}(2022)\citenamefont {Xu} \emph
  {et~al.}}]{Xu2022Prl}%
  \BibitemOpen
  \bibfield  {author} {\bibinfo {author} {\bibfnamefont {K.}~\bibnamefont {Xu}}
  \emph {et~al.},\ }\bibfield  {title} {\enquote {\bibinfo {title}
  {Metrological characterization of non-{G}aussian entangled states of
  superconducting qubits},}\ }\href {\doibase 10.1103/PhysRevLett.128.150501}
  {\bibfield  {journal} {\bibinfo  {journal} {Phys. Rev. Lett.}\ }\textbf
  {\bibinfo {volume} {128}},\ \bibinfo {pages} {150501} (\bibinfo {year}
  {2022})}\BibitemShut {NoStop}%
\bibitem [{\citenamefont {Zhang}\ \emph {et~al.}(2023)\citenamefont {Zhang},
  \citenamefont {Lu}, \citenamefont {Liu}, \citenamefont {Ding},\ and\
  \citenamefont {Wang}}]{Zhang2023Pra}%
  \BibitemOpen
  \bibfield  {author} {\bibinfo {author} {\bibfnamefont {X.}~\bibnamefont
  {Zhang}}, \bibinfo {author} {\bibfnamefont {X.-M.}\ \bibnamefont {Lu}},
  \bibinfo {author} {\bibfnamefont {J.}~\bibnamefont {Liu}}, \bibinfo {author}
  {\bibfnamefont {W.}~\bibnamefont {Ding}}, \ and\ \bibinfo {author}
  {\bibfnamefont {X.}~\bibnamefont {Wang}},\ }\bibfield  {title} {\enquote
  {\bibinfo {title} {Direct measurement of quantum fisher information},}\
  }\href {\doibase 10.1103/PhysRevA.107.012414} {\bibfield  {journal} {\bibinfo
   {journal} {Phys. Rev. A}\ }\textbf {\bibinfo {volume} {107}},\ \bibinfo
  {pages} {012414} (\bibinfo {year} {2023})}\BibitemShut {NoStop}%
\bibitem [{Not()}]{Note_eq16}%
  \BibitemOpen
  \href@noop {} {}\bibinfo {note} {If the transformed probe state
  $\rho_{S}(x)=|\psi(x)\rangle\langle\psi(x)|$ is pure, the QFI can be defined
  as
  $F_{Q}[\rho_{S}(x)]=4\left[\langle\partial_{x}\psi(x)|\partial_{x}\psi(x)\rangle-|\langle\psi(x)|\partial_{x}\psi(x)\rangle|\right]$,
  where $\partial_{x}=\partial/\partial x$ denotes the partial derivative with
  respect to $x$.}\BibitemShut {Stop}%
\bibitem [{\citenamefont {Escher}\ \emph {et~al.}(2012)\citenamefont {Escher},
  \citenamefont {Davidovich}, \citenamefont {Zagury},\ and\ \citenamefont
  {de~Matos~Filho}}]{Escher2012Prl}%
  \BibitemOpen
  \bibfield  {author} {\bibinfo {author} {\bibfnamefont {B.~M.}\ \bibnamefont
  {Escher}}, \bibinfo {author} {\bibfnamefont {L.}~\bibnamefont {Davidovich}},
  \bibinfo {author} {\bibfnamefont {N.}~\bibnamefont {Zagury}}, \ and\ \bibinfo
  {author} {\bibfnamefont {R.~L.}\ \bibnamefont {de~Matos~Filho}},\ }\bibfield
  {title} {\enquote {\bibinfo {title} {Quantum metrological limits via a
  variational approach},}\ }\href {\doibase 10.1103/PhysRevLett.109.190404}
  {\bibfield  {journal} {\bibinfo  {journal} {Phys. Rev. Lett.}\ }\textbf
  {\bibinfo {volume} {109}},\ \bibinfo {pages} {190404} (\bibinfo {year}
  {2012})}\BibitemShut {NoStop}%
\bibitem [{not()}]{note_eq17}%
  \BibitemOpen
  \href@noop {} {}\bibinfo {note} {The introduction of the unitary operator
  $U_{E}$ is to erase all nonredundant information about a variational
  parameter that has been leaked from system space into the full
  space.}\BibitemShut {Stop}%
\bibitem [{\citenamefont {Boyd}\ and\ \citenamefont
  {Vandenberghe}(2009)}]{Boyd2009book}%
  \BibitemOpen
  \bibfield  {author} {\bibinfo {author} {\bibfnamefont {S.}~\bibnamefont
  {Boyd}}\ and\ \bibinfo {author} {\bibfnamefont {L.}~\bibnamefont
  {Vandenberghe}},\ }\href {https://web.stanford.edu/~boyd/cvxbook/} {\emph
  {\bibinfo {title} {Convex Optimization}}}\ (\bibinfo  {publisher} {Cambridge
  University},\ \bibinfo {address} {Cambridge, England},\ \bibinfo {year}
  {2009})\BibitemShut {NoStop}%
\bibitem [{\citenamefont {Pang}\ and\ \citenamefont
  {Jordan}(2017)}]{Pang2017Nc}%
  \BibitemOpen
  \bibfield  {author} {\bibinfo {author} {\bibfnamefont {S.}~\bibnamefont
  {Pang}}\ and\ \bibinfo {author} {\bibfnamefont {A.~N.}\ \bibnamefont
  {Jordan}},\ }\bibfield  {title} {\enquote {\bibinfo {title} {Optimal adaptive
  control for quantum metrology with time-dependent {H}amiltonians},}\ }\href
  {http://dx.doi.org/10.1038/ncomms14695} {\bibfield  {journal} {\bibinfo
  {journal} {Nat. Commun.}\ }\textbf {\bibinfo {volume} {8}},\ \bibinfo {pages}
  {14695} (\bibinfo {year} {2017})}\BibitemShut {NoStop}%
\end{thebibliography}%

\end{document}